\documentclass[aps,twocolumn,amsmath,amssymb,preprintnumbers]{revtex4}
\usepackage{amsmath} \usepackage{amsfonts} \usepackage{amssymb}
\usepackage[utf8]{inputenc}
\usepackage{titlesec}
\usepackage{bbm}
\usepackage{epsfig}
\usepackage{graphics}
\usepackage{graphicx}
\usepackage{lscape}
\newcommand{\be}{\begin{equation}}
\newcommand{\ee}{\end{equation}}
\newcommand{\ba}{\begin{eqnarray}}
\newcommand{\ea}{\end{eqnarray}}
\newcommand{\nn}{\nonumber}
\newcommand{\kl}{\langle}
\newcommand{\kr}{\rangle}

\newcommand{\mn}{_{\mu\nu}}
\newcommand{\pb}{\partial b}

\newcommand{\pJ}{\partial J}
\newcommand{\G}{\bar\Gamma}
\newcommand{\tr}{{\rm tr}}
\newcommand{\pg}{\partial g }

\newcommand{\D}{{\cal D}}
\newcommand{\phg}{\partial \hat g}

\titleformat{\subsection}[block]{\normalfont\bfseries}{\thesubsection.}{1ex}{}
\titlespacing{\subsection}{0pt}{10pt}{1pt}[0pt]
\titleformat*{\section}{\large\bfseries}
\renewcommand{\thesubsection}{\arabic{subsection}}

\begin{document}

\title[ ]{Gauge symmetry from decoupling}

\author{C. Wetterich}
\affiliation{Institut  f\"ur Theoretische Physik\\
Universit\"at Heidelberg\\
Philosophenweg 16, D-69120 Heidelberg}

\begin{abstract}
Gauge symmetries emerge from a redundant description of the effective action for light degrees of freedom after the decoupling of heavy modes. This redundant description avoids the use of explicit constraints in configuration space. For non-linear constraints the gauge symmetries are non-linear. In a quantum field theory setting the gauge symmetries are local and can describe Yang-Mills theories or quantum gravity. We formulate gauge invariant fields that correspond to the non-linear light degrees of freedom. In the context of functional renormalization gauge symmetries can emerge if the flow generates or preserves large mass-like terms for the heavy degrees of freedom. They correspond to a particular form of gauge fixing terms in quantum field theories.

\end{abstract}

\maketitle

\section{Introduction}

Gauge symmetries characterize the fundamental interactions - strong, electroweak and gravitational. Where do they come from? Their most important property is to protect gauge bosons and the graviton from being massive particles, at least perturbatively and in the absence of spontaneous symmetry breaking. From the viewpoint of a microscopic theory that features a large mass scale as the Planck mass, this property guarantees non-trivial long distance physics at length scales much larger than the Planck length. From the perspective of functional flow of the effective action gauge symmetries permit ``life after the Planck mass'' - not all particles get ``heavy masses'' $\sim M$ such that the flow would effectively stop. 

This raises a simple question \cite{Nil,INT,FT,LW,MS,PS,BSS,BBD,Wi,CWSL,CWLL}: Can gauge symmetries emerge macroscopically from the flow for a microscopic setting without gauge symmetries? If yes, gauge symmetries become a property of the infrared (IR) behavior, as seen from some short distance scale $M^{-1}$. The IR-flow should be attracted towards a partial fixed point which realizes gauge symmetry.

For global symmetries such a behavior is well known in a generalized Wilsonian setting of functional flow equations. An enhanced symmetry always constitutes a partial fixed point if the flow equation is compatible with this symmetry. Once the effective action exhibits a symmetry exactly, the flow will not move away from the symmetry. This defines the partial fixed point property. Small deviations from the fixed point may either grow (IR-unstable) with the flow towards the infrared, or they may decrease (IR-stable). A macroscopic global symmetry can be generated by the flow in a more general setting if the partial fixed point is IR-stable.

The situation for local gauge symmetries differs from the case of global symmetries. An exact local symmetry eliminates degrees of freedom which no longer belong to the spectrum of physical excitations. A continuous approach to a local symmetry needs therefore to eliminate continuously these additional degrees of freedom. The most straightforward way how this can be achieved is the generation (or preservation) of a large mass-like term for the additional degrees of freedom. The approach to local symmetry is then the process of decoupling of the ``heavy degrees of freedom''. This should realize local gauge symmetry for the description of the remaining light degrees of freedom. 

A different point of view may state that gauge symmetries as diffeomorphism symmetry are fundamental, with no need or use of having them emerge from a more general setting. It is, however, difficult to realize diffeomorphism invariance in a regularized quantum field theory, in particular if the setting is discrete. In a discrete formulation it is possible to impose lattice diffeomorphism symmetry \cite{CWLD}. This symmetry is, however, not as strong as the continuous diffeomorphism invariance. The latter should be realized in the continuum limit. One is back to the question of emergence of local gauge symmetries from the flow towards the infrared. 

The present paper explores the possibility that local gauge symmetries emerge from the decoupling of heavy degrees of freedom. We first investigate in sect. \ref{Gauge symmetry from decoupling} the most simple case of only one light and one heavy variable with a linear gauge symmetry. Heavy degrees of freedom are defined by the presence of a large quadratic term in the action (``heavy mass''), while for the light degrees of freedom no such term occurs. Correspondingly, we divide the sources in the functional integral into ``physical sources'' $J$ for the light degrees of freedom, and complementary sources $H$ for the heavy degrees of freedom. 

More generally, the physical sources obey a constraint formulated with an appropriate projector $P$, $P^TJ=J$. Correspondingly, the light degrees of freedom $\hat g$ are constrained variables. One defines the effective action $\G(\hat g)$ for the light degrees of freedom by simply setting the heavy degrees of freedom to zero. Omitting the constraint for $\hat g$ and extending $\G(g)$ to unconstrained variables $g$ realizes the gauge symmetry. It corresponds to a redundant description, since $\G$ only depends on $\hat g$ despite its formal dependence on general variables $g$. More precisely, this procedure involves the map $g\to \hat g(g)$ which associates to each general variable $g$ a constrained variable $\hat g(g)$, with $\G(g)=\G\big(\hat g(g)\big)$. The gauge transformations acting on $g$ are the transformations that leave $\hat g(g)$ invariant. 

The constraint on the physical sources should correspond to the covariant conservation of currents for Yang  Mills theories or to the covariantly conserved energy momentum tensor for quantum gravity. These constraints involve the macroscopic gauge fields or metric. We therefore consider in sect. \ref{Non-linear gauge symmetry} the case of a ``field dependent projector'' $P(g)$. Such a field dependence of the projector renders the gauge transformations non-linear, as characteristic for Yang-Mills theories or gravity. The explicit construction of the map $g\to \hat g(g)$ leads to the concept of gauge invariant variables $\hat g(g)$ that we discuss in sect. \ref{Physical variable}. These physical variables $\hat g(g)$ obey differential constraints
\be\label{7Aa}
P(\hat g)\frac{\partial\hat g}{\partial g}=\frac{\partial \hat g}{\partial g}P(g)=\frac{\partial \hat g}{\partial g}.
\ee
For Yang-Mills theories or gravity they will be generalized to gauge invariant fields $\hat A_\mu(x)$ or metrics $\hat g_{\mu\nu}(x)$. The gauge invariant variables $\hat g(g)$ correspond to trajectories in field space rather than being defined globally. Their specification involves ``initial values''. 

We discuss the macrophysical gauge invariant effective action and its properties in sect. \ref{Macroscopic gauge symmetry}. In sect. \ref{Multi-component variables and fields} we generalize our setting from the simple two variable case to $N$ variables. Quantum field theories obtain in the limit $N\to \infty$, as briefly outlined in sect. \ref{Quantum field theories}. The particular gauge symmetries of Yang-Mills theories or gravity specify  the projector $P$ and therefore the form of the effective gauge fixing term for the heavy modes. Within functional renormalization for the effective average action \cite{CWFE} this form appears in the formalism of gauge invariant flow equations involving a single gauge field \cite{CWFLOW}.
 Conclusions are presented in sect. \ref{Conclusions}.

\section{Gauge symmetry for light mode from decoupling of heavy mode}
\label{Gauge symmetry from decoupling}

We will interpret the emergence of a gauge invariant effective action in terms of the decoupling of ``heavy modes" or ``heavy degrees of freedom". Gauge symmetry arises as a redundant description for the ``light modes". Here the notion of ``heavy" and ``light" is associated to the presence or absence of a large quadratic term in the action, similar to the mass term for particles in quantum field theory. (Actually, only the relative size of the quadratic terms for light and heavy degrees of freedom matters.) In this section we describe the setting in its simplest version, with one heavy mode $c$ and one light mode $b$, and a linear gauge symmetry.

The emergence of gauge symmetry can be sketched as follows: The effective action $\Gamma[ g ]=\Gamma(b,c)$ has a large quadratic term in $c$ and no such term for $b$. We can construct an effective theory for the light mode $b$ by simply setting $c=0$, 
\be\label{36A}
\bar\Gamma(b)=\Gamma(b,c=0).
\ee
For $\bar \Gamma(b)$ any dependence on the heavy field $c$ is eliminated. In particular, the large quadratic term for $c$ (which corresponds to the gauge fixing term) is no longer present. We may now formally reintroduce $c$ by using $g=(b,c)$, while maintaining the effective action $\bar \Gamma$. The formal appearance of $g$ in $\bar\Gamma(g)$ is redundant, since $\bar\Gamma$ actually only depends on the light mode $b$. This redundancy is reflected by the  invariance
\be\label{36B}
\Gamma\big (g+(0,c)\big)=\Gamma(g).
\ee
The shift symmetry under $g\to g+(0,c)$ corresponds to the local gauge symmetry once $b$ and $c$ are promoted to fields. Thus gauge symmetry arises as a formal invariance in a redundant description, which tells that the only physical degree of freedom in the effective theory is $b$, despite the formal appearance of $g=(b,c)$.

We demonstrate the basic idea here in its simplest form, with only one light variable $b$ and one heavy variable $c$. The coefficient of the quadratic term for $c$ is taken to be constant (field independent). In this form the usefulness of a gauge invariant formulation for the light mode is not yet clear - this will become more apparent in the next section where the projections on the light and heavy modes in field space are more complex. Nevertheless, many key features of our setting are already visible in this simplest example.

\subsection{Field independent source constraint}

We first want to understand the circumstances under which our setting for a gauge invariant effective action $\bar \Gamma$ arises from the generating functional $W$ for connected correlation functions. Consider a function $W$ that depends on two sources $J$ and $H$,

\be\label{B1}
W=\frac\rho2 J^2+\frac\alpha2 H^2+\gamma JH,
\ee
where $\rho$ may depend on $J$. The sources $J$ and $H$ couple to the light and heavy degrees of freedom, respectively. The source constraint projecting on the ``physical source'' for the light degree of freedom simply reads $H=0$. The ``macroscopic fields'' associated to $J$ and $H$ are
\ba\label{B2}
b&=&\frac{\partial W}{\partial J}=\rho J+\frac12\frac{\partial \rho}{\partial J}J^2+\gamma H,\nn\\
c&=&\frac{\partial W}{\partial H}=\alpha H+\gamma J.
\ea
Instead of fields we deal here, however, with simple variables $b,c,J,H$ and simple functions $W$ and $\Gamma$.

The effective action is defined by the Legendre transform
\ba\label{B3}
\Gamma&=&-W+J b+Hc=\frac\rho2 J^2+\frac{1}{2}\frac{\partial\rho}{\partial J}J^3+\frac\alpha2 H^2+\gamma JH\nn\\
&=&\frac\rho2 J^2+\frac12 \frac{\partial \rho}{\partial J}J^3+\frac{\gamma^2}{2\alpha}J^2-\frac\gamma\alpha Jc+\frac{c^2}{2\alpha},
\ea
where the last identity uses
\be\label{B3A}
H=\frac c\alpha-\frac\gamma\alpha J.
\ee
Similarly, $J$ is considered as a function of $b$ and $c$, given by the solution of eq. \eqref{B2}. We will not need its explicit form. One has the usual identities
\be\label{8A}
\frac{\partial\Gamma}{\partial b}=J~,~\frac{\partial\Gamma}{\partial c}=H.
\ee

In the limit $\gamma/\alpha\to 0$ the effective action decomposes into two separate pieces
\be\label{B4}
\Gamma=\bar \Gamma(b)+\Gamma_{gf}(c)~,~\Gamma_{gf}=\frac{c^2}{2\alpha},
\ee
with $\bar \Gamma(b)$ the Legendre transform of $\bar W(J)=(\rho/2)J^2$. For $\gamma\neq 0$ we may define $\bar  \Gamma(b)=\Gamma(b,c=0)$ and observe for $\bar{\Gamma} \left( b \right)$ an additional term $\sim\gamma^2$.

Consider now the two-component vectors $g=(g_1,g_2)=(b,c)$ and $L=(L_1,L_2)=(J,H)$, such that 
\be\label{9A}
\frac{\partial W}{\partial L_i}=g_i~,~\frac{\partial \Gamma}{\partial g_i}=L_i.
\ee
With $b(g)=g_1$ the function $\bar\Gamma(b)$ can be written as $\bar\Gamma(g)=\G\big(b(g)\big)$, which actually only depends on the first component $g_1=b$,  such that 
\be\label{B5}
\frac{\partial\bar\Gamma}{\partial g_1}=L_1~,~\frac{\partial\bar\Gamma}{\partial g_2}=0.
\ee
This realizes the property that the field equations only involve the physical source. The second equation \eqref{B5} implies a ``gauge symmetry'' under the infinitesimal transformation
\be\label{B6}
\delta_\xi g_i=-\xi\delta_{i2}~,~\delta_\xi\bar \Gamma=\frac{\partial\bar \Gamma}{\partial g_i}\delta_\xi g_i=0.
\ee

For $W$ we have for $H=0$ the identity
\be\label{B7}
\frac{\partial W}{\partial L_1}_{|H=0}=g_1=b,
\ee
where $b=b(J,H=0)$. The second component,
\be\label{B8}
\frac{\partial W}{\partial L_2}_{|H=0}=\gamma J=\gamma\frac{\partial \Gamma}{\partial b},
\ee
vanishes only for $\gamma\to 0$. 

The projection on the light field or physical variable takes a simple form
\be\label{B9}
\hat g=P\binom bc=\binom b0 ~,~P=\binom{1,0}{0,0}~,~P^2=P.
\ee
Similarly, the projection on the ``physical source'' $J$ obeys 
\be\label{12A}
{J \choose 0} =Pg.
\ee
This realizes a linear gauge symmetry where the projector is filed independent, similar to abelian local gauge theories as pure QED.

\subsection{Physical effective action}

Our example shows that gauge symmetry can be realized rather trivially by defining $\G(g)=\Gamma(g,c=0)$. We will require further that $\G$ is the ``physical effective action'', realized for physical sources, e.g. $H=0$. Beyond the formal gauge invariance we will impose conditions such that $\G(g)$ indeed describes the effective action for the light degrees of freedom,corresponding to a restriction to physical sources. The first condition simply states that for a restriction to physical sources the macroscopic variable for the heavy degree of freedom should vanish
\be\label{14A}
c(J,H=0)=0.
\ee
Otherwise the effective action $\G(b)=\Gamma(b,c=0)$ describes the system only for $H\neq 0$. For the second condition we require that the second derivative of $\G$ yields the two point function for the light modes by inversion
\be\label{14B}
\frac{\partial^2\G(b)}{\partial b^2}\frac{\partial^2 W}{\partial J^2}=1.
\ee
In terms of  the projector $P$ eq. \eqref{14B} can be written as
\be\label{14C}
\G^{(2)}_P W^{(2)}_P=P,
\ee
where 
\be\label{14D}
\G^{(2)}_P=P\G^{(2)} P~,~W^{(2)}_P=PW^{(2)}P,
\ee
and $\G^{(2)},~W^{(2)}$ denote the matrices of second derivatives, e.g.
\be\label{14E}
\G^{(2)}_{ij}=\frac{\partial^2\G}{\partial g_i\partial g_j}.
\ee
Again, we require that eq. \eqref{14B} or \eqref{14C} holds for physical sources, i.e. $H=0$. 

From eq. \eqref{B8} and $c=\partial W/\partial L_2$ one immediately concludes that the first condition \eqref{14A} only holds for $\gamma=0$. The projected second derivatives obey
\ba\label{B10}
W^{(2)}_P&=&PW^{(2)}P=
\left(\begin{array}{ccc}
\frac{\partial b }{\partial J}_{|H}&,&0\\  0&,&0
\end{array}\right),\nn\\
\Gamma^{(2)}_P&=&P\Gamma^{(2)}P=
\left(
\begin{array}{ccc}
\frac{\partial^2\Gamma}{\partial b^2}_{|c}&,&0\\
0&,&0
\end{array}
\right).
\ea
One infers the matrix identities
\ba\label{B11}
\Gamma^{(2)}_PW^{(2)}_P&=&\frac{\partial^2\Gamma}{\partial b^2}_{|c}\frac{\pb}{\pJ}_{|H}P\nn\\
&=&\frac{\pJ}{\pb}_{|c}\frac{\pb}{\pJ}_{|H}P,
\ea
and
\be\label{B12}
(1-P)\Gamma^{(2)}W^{(2)}_P=
\frac{\pJ}{\partial c}_{|b}\frac{\pb}{\pJ}_{|H}
\binom{0,0}{1,0}.
\ee
We want to evaluate these relations for $H=0$, corresponding to $c=\gamma J$. In general, $(1-P)\Gamma^{(2)}W_P^{(2)}$ does not vanish and $\Gamma^{(2)}_PW^{(2)}_P$ does not necessarily equal $P$. For $\gamma=0$, however, the source $J$ becomes only a function of $b$ for arbitrary $H$, such that 
\be\label{B13}
\Gamma^{(2)}_PW^{(2)}_P=P.
\ee
Realizing that $\G^{(2)}_P=\Gamma^{(2)}_P$, eq. \eqref{14C} is obeyed for $\gamma=0$. In this case also the r.h.s of eq. \eqref{B12} vanishes due to $\partial J/\partial c=0$, implying 
\be\label{58A}
\Gamma^{(2)}W^{(2)}_P=P.
\ee
We conclude that the two conditions \eqref{14A}, \eqref{14C} amount to the condition $\gamma=0$.

\subsection{Generating function for connected $n$-point}

~{\bf functions}

So far we have considered a given $W$ and discussed its relation to the effective action $\Gamma$. We next want to realize $W=\ln Z$ as the generating function for connected correlation functions in a microscopic formulation. The usual functional integral for the partition function $Z$ is here represented as a simple integral. We will see that our scenario can be realized in the usual setting with gauge fixing, but only provided that a particular form of the gauge fixing is chosen.

Let us define $W(J,H)$ by the integral
\be\label{B14}
W=\ln \int db' dc' \exp \big\{ -S(b',c')-S_{gf}(b',c')+Jb'+Hc'\big\}.
\ee
Here $S$ stands for the microscopic or classical action and $S_{gf}$ is ``a gauge fixing term'' that we first take as
\be\label{B15}
S_{gf}=\frac{1}{2\alpha}c'^{2}.
\ee
The gauge fixing term plays the role of the ``heavy mass'' for the heavy degree of freedom $c'$, and the decoupling limit will correspond to $\alpha\to 0$.

Derivatives of $W$ with respect to $J$ and $H$ yield the connected correlation functions for $b'$ and $c'$. In particular, one has
\be\label{B15A}
\frac{\partial W}{\pJ}=\kl b'\kr =b~,~\frac{\partial W}{\partial H}=\kl c'\kr =c,
\ee
and
\be\label{B15B}
\frac{\partial^2 W}{\partial J^2}=\kl b'^2\kr -b^2.
\ee
If $S$ is gauge invariant under the infinitesimal transformation $\delta_\xi c'=-\xi$, $\delta_\xi b'=0$, it is independent of $c'$. In this case the integral \eqref{B14} yields
\be\label{B16}
W=\bar W(J)+\frac\alpha2 H^2,
\ee
and therefore indeed $\gamma=0$. This demonstrates how our setting can be realized in the most simple form.

Already at this stage we arrive at one of the important conclusions of this paper. It is crucial that the gauge fixing is quadratic in the field $c'$ and does not involve any linear term. Consider a different gauge fixing term 
\be\label{B17}
S_{gf}=\frac{1}{2\alpha}(c'-\epsilon b')^2.
\ee
Performing the Gaussian integration over $c'$ one arrives now at
\ba\label{B18}
W&=&\frac\alpha2 H^2+c+\ln \int db' \exp \big\{-S(b')+(J+\epsilon H)b'\big\}\nn\\
&=&\bar W(J+\epsilon H)+\frac\alpha 2 H^2,
\ea
where we have taken a gauge invariant microscopic action $S \left( b' \right)$.
For $\bar W(J)=(\rho/2)J^2$ this produces a term linear in $H$,
\ba\label{B19}
\bar W&=&\frac{\rho(J+\epsilon H)}{2}(J^2+2\epsilon JH+\epsilon^2 H^2)\nn\\
&=&\bar W(J)+\gamma(J)HJ+0(H^2),
\ea
where
\be\label{B20}
\gamma=\epsilon\left[\rho(J)+\frac J2\frac{\partial \rho(J)}{\partial J}\right].
\ee
The conditions \eqref{14A}, \eqref{14C} for a physical effective action are therefore not realized for an arbitrary form of the gauge fixing!

In the presence of a non-zero $\gamma$ one may still define a gauge invariant effective action $\bar \Gamma$ using
\be\label{B21}
\bar \Gamma(b)=\Gamma(b,c=0)~,~\frac{\partial\bar \Gamma}{\pb}=J.
\ee
Making the trivial extension $\bar \Gamma(g)=\bar \Gamma(g_1,g_2)=\bar \Gamma(g_1)$, one arrives at a gauge invariant effective action $\bar \Gamma(g)$. The definition \eqref{B21} corresponds to a ``wrong expansion point'' since $H$ differs from zero for $c=0$, but one may not care and be satisfied with $\partial\G/\partial b=J$. What goes wrong, however, is the connection between the second derivative of $\bar \Gamma$ and the connected correlation function $W^{(2)}_P$. For the projected second derivative,
\be\label{B22}
\bar \Gamma^{(2)}_P=P\bar \Gamma ^{(2)} P=\frac{\partial^2\bar \Gamma}{\pb^2}P,
\ee
the relation $\bar\Gamma^{(2)}_PW^{(2)}_{P|H=0}=P$ no longer holds, such that the correlation function for the light modes for $H=0$ cannot be extracted from $\G$. 

The origin of this problem is apparent on the level of $\Gamma$ where we can employ $\Gamma^{(2)}W^{(2)}=1$. For
\be\label{B23}
W^{(2)}{_{|H=0}}=\binom{\tilde\rho~,~ \gamma}{\gamma~,~\alpha}
\ee
one has
\be\label{B24}
\Gamma^{(2)}=\frac{1}{\tilde \rho\alpha-\gamma^2}
\left(\begin{array}{ccc}
\alpha&,&-\gamma\\ -\gamma&,&\tilde \rho
\end{array}\right),
\ee
and therefore
\be\label{B25}
\Gamma^{(2)}_PW^{(2)}_P=\frac{\tilde\rho\alpha}{\tilde \rho\alpha-\gamma^2}P=
\left(1-\frac{\gamma^2}{\tilde \rho\alpha}\right)^{-1}P.
\ee
The r.h.s. equals the projector only for $\gamma=0$. Similarly, for $\bar\Gamma(b)$ defined by eq. \eqref{B21} one has
\be\label{B26}
\frac{\partial^2\bar\Gamma}{\pb^2}=\frac{\partial ^2\Gamma}{\pb^2}_{|c=0}=\frac{\alpha}{\tilde \rho\alpha-\gamma^2},
\ee
reproducing eq. \eqref{B25}. We conclude that $W^{(2)}_P$ is no longer the inverse of $\bar \Gamma^{(2)}_P$ in the projected space. This relation is crucial, however, in order to compute correlation functions from $\bar\Gamma$ or to formulate an exact flow equation.

\subsection{Gauge invariant effective action without \\ \hspace*{0,5cm} microscopic gauge invariance}

\medskip
Consider next the case where the action $S(b',c')$ in eq. \eqref{B14} depends on $c'$,
\be\label{B27}
S(b',c')=\bar S(b')+r_1(b')c'+\frac12 r_2(b') c'^2+\dots
\ee

\noindent
We will show that our scenario can be realized even for a non-gauge invariant microscopic action (\ref{B27}), provided we take $\alpha \to 0$ in the gauge fixing term \eqref{B15}.

We first keep only the term linear in $c'$. (Higher order terms can be combined with the gauge fixing term. We will be interested in the limit $\alpha\to 0$ for which they can be neglected.) Performing the Gaussian integral over $c'$ yields
\be\label{B28}
W=\frac\alpha2 H^2+\ln\int db'\exp
\big\{-\bar S(b')+\frac\alpha2 r^2_1-\alpha r_1H+b'J\big\}.
\ee
This produces a linear term in $H$ corresponding to $\gamma\neq 0$,
\be\label{B29}
\frac{\partial W}{\partial H}_{|H=0}=-\alpha\kl r_1\kr_{|J},
\ee
where the expectation value $\kl r_1\kr$ is evaluated with the action
\be\label{B29A}
S'=\bar S-\frac{\alpha r^2_1}{2}
\ee
and in presence of the source $J$.

In the limit $\alpha\to 0$ the influence of the term linear in $H$ becomes negligible. In this limit one has $S'=\bar S$. In particular, for $r_1=rb'$ with constant $r$ one has
\be\label{B30}
\gamma=-\frac{\alpha rb}{J}=-\alpha r\left(\rho+\frac12\frac{\partial\rho}{\pJ}J\right).
\ee
Corrections $\sim \gamma^2/\tilde \rho\alpha$ in eq.\eqref{B25} vanish $\sim \alpha\tilde \rho r^2$ for $\alpha\to 0$. Furthermore, the term $\sim r_2 c'^2$ in eq.\eqref{B27} can be neglected as compared to $c'^2/2\alpha$. For $\alpha\to 0$ the term $\exp (-c'^2/2\alpha)$ becomes $\delta(c')$, up to an irrelevant constant factor. Therefore also all higher order terms in the expansion \eqref{B27} become negligible. We conclude that our setting with a gauge invariant effective action $\bar\Gamma[g]$ and the properties \eqref{14A}, \eqref{14C} is realized for $\alpha\to 0$ even if the microscopic action is not gauge invariant, e.g. $r_1\neq 0$. The gauge symmetry violating terms are ``projected out'' by the gauge fixing term for $\alpha\to 0$. 

Let us describe this issue in more detail. Up to first order in $\alpha$ one finds
\be\label{B31}
W=\bar W(J)+\frac\alpha2 H^2+\alpha T(J)H,
\ee
with
\ba\label{B32}
\bar W(J)&=&\ln \bar Z(J),\nn\\
\bar Z(J)&=&\int db'\exp \big\{-S'(b')+Jb'\big\},
\ea
and
\be\label{B33}
T(J)=-\kl r_1\kr=-\bar Z^{-1}\int db' r_1(b')\exp \big\{-S'(b')+Jb'\big\}.
\ee

\noindent
Thus $W$ has the same form as eq.\eqref{B16} up to the ``linear term'' $\alpha TH$. The linear term does not affect $W_P^{\left( 2 \right)}$.

In the presence of this linear term the expectation value of $c'$ no longer vanishes for $H=0$ and nonzero small $\alpha$,
\be\label{B34}
\kl c'\kr =c=\alpha \big(H+T(J)\big).
\ee
Inverting this equation, together with

\begin{align}
\left\langle b' \right\rangle = b = \frac{\partial \bar{W}}{\partial J} + \alpha H \frac{\partial T}{\partial J},
\end{align}

\noindent
one obtains the effective action

\begin{align}
\Gamma \left( b, c \right) = \bar{\Gamma} \left( b \right) + \frac{\alpha}{2} H^2 + \alpha H J \frac{\partial T}{\partial J},
\end{align}

\noindent
with

\begin{align}
\bar{\Gamma} \left( b \right) = \frac{\partial \bar{W}}{\partial J} J - \bar{W}
\end{align}

\noindent
the Legendre transform of $\bar{W} \left[ J \right]$. For $\alpha \to 0$ one has 

\begin{align}
\frac{\partial \bar{W}}{\partial J} = b, \quad \frac{\partial \bar{\Gamma}}{\partial b} = J.
\end{align}

The second derivative of $\Gamma$ reads

\begin{align}
\Gamma^{\left( 2 \right)} = \frac{1}{\bar{W}^{\left( 2 \right)} - \alpha \left( \frac{\partial T}{\partial J} \right) ^2} \left( \begin{matrix}
1 & ~,~ -\frac{\partial T}{\partial J} \\[6pt]
-\frac{\partial T}{\partial J} & ~,~ \frac{1}{\alpha} \bar{W}^{\left( 2 \right)} 
\end{matrix} \right),
\end{align}

\noindent
such that for $\alpha \to 0 $ the relation 

\begin{align}
\Gamma_P^{\left( 2 \right)} W_P^{\left( 2 \right)} = 1
\end{align}

\noindent
is obeyed. Extending $\bar{\Gamma}\left( b \right)$ to $\bar{\Gamma}\left( g \right)$ we realize eq.$(\ref{B5})$ and gauge invariance of $\bar{\Gamma}$ according to eq.$(\ref{B6})$. The projected second functional derivative $\Gamma_P^{\left( 2 \right)}$ can be computed from $\bar{\Gamma} \left( g \right)$.

Since for $\alpha \to 0 $ one has $c \sim \alpha$, the influence of the higher order terms in the expression \eqref{B27} is suppressed by higher powers of $\alpha$. For example, one has

\begin{align}\label{38G}
\left\langle c'^2 \right\rangle = c^2 + \frac{\partial^2 W}{\partial H^2} = \alpha+\alpha^2  \left( H + T \right)^2.
\end{align}

\noindent
This type of expression appears (multiplied with $r_2\left( b' \right)$) if we insert the expression $(\ref{B27})$ into the integral over $c'$. For $\alpha \to 0$ such terms can be neglected. We conclude that for $\alpha\to 0$ the functions $\bar W(J)$ and $\bar \Gamma(b)$ can be computed by replacing $S\to \bar S(b')$ in the integral (\ref{B14}). They remain related by a Legendre transform.

\subsection{Gauge symmetry from decoupling}

This simple finding has an important consequence: the limit $\alpha\to 0$ ``projects out'' any gauge symmetry violating term in the ``microscopic action'' $S$. We do not need to start with a gauge invariant microscopic action $S$ in order to obtain a gauge invariant effective action $\bar\Gamma$. In particular, we may add an infrared cutoff violating gauge invariance. It may introduce a term $\sim r_1 c'$, but the influence of this gauge symmetry breaking vanishes if we choose a gauge fixing with $\alpha\to 0$.

For the present example this has a very simple interpretation. For $\alpha\to 0$ the variable $c'$ corresponds to a ``heavy degree of freedom'', while $b'$ can be viewed as a ``light degree of freedom''. In the limit $\alpha\to 0$ the heavy degree of freedom ``decouples'' and the effective action for the light degree of freedom no longer feels the influence of the heavy degree of freedom. Gauge invariance expresses the fact that after omission of the heavy field the effective action for the light field $\bar \Gamma$ only depends on $b$. The generating functions $\bar W(J)$ and $\bar \Gamma(b)$ are related by a Legendre transform. What is important in this setting is not the gauge invariance of the microscopic action $S$, but rather the correct form of the quadratic term in the heavy fields $S_{gf}$. In the decoupling limit $\alpha\to 0$ it is the form of $S_{gf}$ that determines the precise gauge symmetry of the effective action for the light modes. 

We also note that for $\alpha\to 0$ an inappropriate gauge fixing term \eqref{B17} cannot be ``cured'' by any finite non-zero source $H$. A finite linear term cannot modify the relation $c=\epsilon b$, such that the conditions for a physical effective action do not only follow from $H=0$, but from any finite $H$. 

Our simplest example gives already a glance how gauge symmetry can be realized by the flow to the infrared. It is sufficient that a term $\Gamma_{gf}$ of the form \eqref{B4} is generated, and that $\alpha$ reaches the decoupling limit $\alpha\to 0$. Nothing else is required for the microscopic setting, provided $\Gamma_{gf}$ is the only term in the effective action that diverges for $\alpha\to 0$.

We will see next how these properties generalize to settings more closely related to Yang Mills theories and gravity. The main conceptual difference will be that light and heavy field cannot be defined globally, but only locally in field space. 

\section{Non-linear gauge symmetry}
\label{Non-linear gauge symmetry}

The situation for quantum gravity or non-abelian gauge theories differs in one important aspect from the setting of the previous section. In gravity the covariant conservation of the energy momentum tensor amounts to a constraint on the source that involves the metric. Similarly, the physical sources for non-abelian gauge theories are covariantly conserved currents, such that the constraint involves the macroscopic gauge field. We therefore have to generalize our setting to the case of a constraint on the source that depends on a macroscopic ``field'' $g=(g_1,g_2)$. In this section we will again consider only two variables $g_1$ and $g_2$.  The projector depends now on the macroscopic field $g$ and therefore indirectly on the physical source. Therefore the precise meaning of light and heavy degrees of freedom will be modified. In particular, the ``physical variable'' $\hat g$ for the light degree of freedom can no longer be defined globally but rather obeys a differential constraint.

\subsection{Field dependent projector}

The main new ingredient of this section is the dependence of the projector $P$ on the macroscopic field $g$ and thereby indirectly on the source $L$. The projection property or constraint for the physical source,
\be\label{50A}
J=P(g)L,
\ee
will define the precise gauge symmetry. 

We first consider a symmetric projector
\be\label{CB1}
P=\frac{1}{1+\eta^2}
\left(\begin{array}{rc}
1,&\eta\\ \eta,&\eta^2\end{array}\right)~,~P^T=P,
\ee
that obeys
\be\label{CB2}
P^2=P~,~1-P=\frac{1}{1+\eta^2}
\left(\begin{array}{rc}
\eta^2,&-\eta\\-\eta,&1
\end{array}\right).
\ee
Here $\eta$ is a function of $g_1$ and $g_2$ such that the projector $P=P(g)$ depends on $g$. 

The physical source $J$ obeys the field dependent constraint
\be\label{CB3}
J=PL=\frac{L_1+\eta L_2}{1+\eta^2}\binom{1}{\eta}~,~PJ=J.
\ee
If $\bar \Gamma(g)$ couples to the physical source,
\be\label{CB4}
\frac{\partial\bar\Gamma}{\partial g_i}=J_i,
\ee
this implies gauge invariance under the infinitesimal transformation
\be\label{CB5}
\delta_\xi g=
-\xi v(g)~,~v(g)=\frac{1}{\sqrt{1+\eta^2}}\binom{-\eta}{~~1}
~,~\delta_\xi\bar\Gamma=0,
\ee
as follows directly from
\be\label{CB6}
\delta_\xi\bar\Gamma=\frac{\partial\bar\Gamma}{\partial g_i}\delta_\xi g_i=J_i\delta_\xi g_i=0.
\ee
The gauge invariance of $\bar\Gamma$ follows from the redundant description, using unconstrained variables $g$ while the source is constrained. We observe that the gauge variation \eqref{CB5} of the macroscopic variable involves $g$ in a non-linear way, as characteristic for the field dependence of gauge transformations in gravity or non-abelian gauge theories.

We may expand $g$ around a given ``expansion point'' or ``background'' $\bar g$
\be\label{CB6a}
g=\bar g+h.
\ee
Infinitesimal fluctuations $h$ can be split into ``physical fluctuations'' $f$,
\be\label{CB8}
f=P(\bar g)h=\frac{h_1+\eta h_2}{1+\eta^2}\binom 1\eta,
\ee
and ``gauge fluctuations'' $a$,
\be\label{CB9}
a=\big(1-P(\bar g)\big)h=\frac{h_2-\eta h_1}{1+\eta^2}\binom{-\eta}{1}~,~h=f+a,
\ee
with
\be\label{CB9A}
f^Tf+a^Ta=h^Th=h^2_1+h^2_2.
\ee

Consider now the gauge transformation of the background field $\bar g$
\be\label{CB10}
\delta_\xi\bar g=-\xi v(\bar g).
\ee
This transformation  can equivalently be accounted for by an infinitesimal change of $a$, leaving $f$ and $\bar g$ invariant,
\be\label{CB11}
\delta_\xi a=
-\xi v(\bar g)
~,~\delta_\xi f=0~,~\delta_\xi\bar g=0.
\ee
An infinitesimal gauge fluctuation $a$ can therefore be viewed as the change of $\bar g$ under gauge transformations, hence the name ``gauge fluctuation''. For finite $f$ and $a$ the gauge transformation of $g$ induces additional terms in $\delta_\xi f$ and $\delta_\xi a$. Due to the dependence of $\eta$ on $g$ a simple global gauge degree of freedom no longer exists. 

The choice of the expansion point $\bar g$ is completely arbitrary - any value of $g$ can be used as expansion point. It makes no sense to split $g$ globally into heavy and light degrees of freedom. However, for each $g$ the physical and gauge ``directions'' are well defined by the projections of an infinitesimal variation $h$. Also the meaning of the projected second derivative $\bar \Gamma^{(2)}_P=P^T\bar\Gamma^{(2)}P$ is well defined. In sect. \ref{Physical variable} we will use the projections \eqref{CB8}, \eqref{CB9} of the fluctuations in order to define the physical variable $\hat g$ by a differential constraint.

\subsection{Generating functions from fluctuation integral}

\medskip
We next investigate how a field dependent constraint on the source is realized within the formulation of a ``functional'' integral. The partition function $Z(L)$ is defined as
\be\label{Z1}
Z=\int d g'_1 dg'_2\exp \big\{-S(g')-\frac{1}{2\alpha}c'^T c'+L^T g'\big\},
\ee
with
\be\label{Z2}
g'= \left( g'_1,g'_2 \right) =\bar g+h' = \bar g+ b'+ c'.
\ee
The fluctuating variables $c'$ and $b'$ are eigenvectors of the projector $P(g)$,
\be\label{Z2a}
P(g) b' = b', \quad \left( 1 - P\left( g \right) \right) c' = c',
\ee
and obey 
\be\label{eq.105}
b' = P \left( g \right) h', \quad c' = \left(1 - P \left( g \right) \right) h'.
\ee
The projector $P \left( g \right)$ depends on the macroscopic field $g = \left( g_1, g_2 \right)$, which is a function of the sources, $g=g(L)$. The precise choice of $g(L)$ will be discussed later. The general source $L$ in eq. \eqref{Z1} can be decomposed into $J$ and $H$, with $J$ the ``physical source'',
\ba\label{Z3}
J=P(g)L~,~H=\big(1-P(g)\big)L.
\ea

The definition of the generating function for connected correlations $W$ involves both $g$ and $\bar g$ in addition to $L$,
\be\label{75A}
W(L,g,\bar g)=\ln Z(L,g,\bar g).
\ee
This occurs since
\be\label{75B}
c'=\big(1-P(g)\big)(g'-\bar g)
\ee
depends on $g$ via the projector and on the ``expansion point'' $\bar g$. We will proceed later to a particular choice of $\bar g(g)$, auch that $W(L,g)=W\big (L,g,\bar g(g)\big)$ depends on $L$ and $g$. Such a choice is understood implicitly in the following and we do not write the dependence on $\bar g$ explicitly. We also will take later the limit $\alpha\to 0$. 

The appearance of the field dependent projector introduces an unfamiliar feature in the definition of the functional integral. The formulation of the partition function $Z$ involves the macroscopic field $g$. More precisely, the argument of the exponential in eq. $(\ref{Z1})$ has no longer a purely linear dependence on the source $L$.  The ``gauge fixing term'' $\sim 1/\alpha$ depends on $c'$. In turn $c'$ depends on $g$ via eq. \eqref{75B} and therefore on the sources, $g=g(L)$. As a consequence, the integral $(\ref{Z1})$ is only defined implicitly. Formally, eq.$(\ref{Z1})$ becomes an integro-differential equation for $Z$ since the r.h.s. depends on $g(L)$. In general, the macroscopic variable $g$ no longer equals the expectation value of the microscopic variable $\left\langle g' \right\rangle$. The relation $g = \left\langle g' \right\rangle$ is not needed, however. 

In the limit $\alpha\to 0$ the ``non-linear formulation'' of the partition function resembles in many aspects a ``linear background formulation'' where the projector is formulated with a fixed background variable $\bar g$ instead of the macroscopic variable $g$. The reason why we formulate the gauge fixing term in terms of the projector $P(g)$ and use a dynamical $\bar g(g)$, rather than employing a fixed $\bar g$ and $P(\bar g)$, arises from the coupling of the ``light variable'' $b'$ to the physical source $J$, which involves $P(g)$ according to eq. \eqref{CB3}. Indeed, with our construction using $P \left( g \right)$ the source term obeys the simple relation
\be\label{Z3a}
L^Tg'=L^T\bar g+J^Tb'+H^T c'.
\ee
This ensures that the light vector $b'$ couples to the physical source, while the heavy vector $c'$ couples to $H$. We do not impose at this point that $S$ is a gauge invariant function of $g'$.

\smallskip

Our implicit construction seems perhaps cumbersome. We will never need, however, to solve the integro-differential equation explicitly. We observe that the quadratic term $\sim c'^{2}/\alpha$ is close to a particular background gauge fixing. The latter would be realized if we use a fixed background field $\bar g$ in the projector, i.e. replacing $P(g)$ by $P(\bar g)$ in eqs. \eqref{Z2a}, \eqref{Z3}. 

We write the generating function
\be\label{109A}
W(L,g)=\ln Z(L,g)=\tilde W(L,g)+L^T\bar g,
\ee
with 
\be\label{109B}
\tilde W(L,g)=\ln \int d g'\exp 
\left\{-S-\frac{1}{2\alpha}c'^Tc'+L^Th'\right\},
\ee
and $h'=g'-g$. One observes the standard relations
\be\label{109C}
\frac{\partial W}{\partial L^i}_{|g}=\kl g'_i\kr=g_i+\kl h'_i\kr~,~
\frac{\partial\tilde W}{\partial L^i}_{|g}=\kl h'_i\kr=h_i.
\ee
Here partial derivatives are taken at fixed $g$ and $\bar g=\bar g(g)$. Both $\kl g'\kr$ and $h$ depend on $g$. 

For fixed $g$ we define $\tilde \Gamma$ as the Legendre transform of $\tilde W$,
\be\label{109D}
\tilde \Gamma(h,g)=-\tilde W(L,g)+L^Th,
\ee
where $L$ is the source associated to $h$ by inverting the second equation \eqref{109C}. One has the usual relations for the first and second derivatives taken at fixed $g$,
\be\label{109E}
\frac{\partial\tilde\Gamma}{\partial h_i}_{|g}=L^i~,~\tilde \Gamma^{(2)}_{|g}\tilde W^{(2)}_{|g}=1.
\ee
Writing
\be\label{109F}
\tilde \Gamma(h,g)=-\tilde W(L,g)-L^Tg+L^T(g+h),
\ee
and observing that the source associated to $h$ is the same as the one associated to $\kl g'\kr=g+h$, one sees that $\tilde \Gamma(h,g)$ can be identified with the Legendre transform of $W(L,g)$ in eq. \eqref{109A}, taken at fixed $g$,
\be\label{109G}
\tilde \Gamma(h,g)=-W(L,g)+L^T(g+h).
\ee
It will be our aim to extract from $\tilde\Gamma(h,g)$ a gauge invariant effective action $\G(g)$. 

In order to proceed we need to specify the relation between the sources $L$ and macroscopic variable $g$, $L=L(g)$, as well as the choice of $\bar g(g)$. Then $h$ is, in principle, computable as a function of $g$. Inserting $h(g)$ in eq. \eqref{109G} the effective action 
\be\label{85A}
\tilde\Gamma(g)=\tilde\Gamma\big(h(g),g\big)
\ee
will only depend on a single variable $g$. We will begin in the next section by identifying $\bar g(g)$ with the ``physical variable'' $\hat g(g)$.

\section{Physical variable}
\label{Physical variable}

The physical variable $\hat g$ plays an important role for the construction of the gauge invariant effective action $\G$. Indeed, the construction of a gauge invariant action $\bar\Gamma(g)$ from the more general effective action $\Gamma(g)$ without gauge invariance relies in a first step on the restriction of the variable $g$ to the ``physical variable'' $\hat g$. The latter obtains by eliminating the heavy degree of freedom, imposing the constraint $\hat c=0$.

\subsection{Physical variable from physical fluctuation}

An arbitrary variable $g$ can be decomposed into a physical variable $\hat g$ and a ``gauge variable'' $\hat c$,
\be\label{V1}
g=\hat g+\hat c.
\ee
We therefore can also write
\be\label{V2}
\hat g=g_{|\hat c=0}.
\ee
The split \eqref{V1} of the macroscopic variable $g$ into physical and gauge variables is a key for our construction. It will find its equivalent on the level of quantum field theory, corresponding to physical and gauge degrees of freedom. We therefore discuss this concept in some detail in our simple two-variable model.

Consider two neighboring variables $g=\bar g$ and $g=\bar g+h$. Infinitesimal fluctuations $h$ can be decomposed into physical fluctuations $f$ and gauge fluctuations $a$ according to eqs. \eqref{CB6a} - \eqref{CB9},
\be\label{V3}
f=P(\bar g)h~,~a=\big(1-P(\bar g)\big)h~,~h=f+a.
\ee
This decomposition of fluctuations will be the basis for the definition of physical and gauge variables and the decomposition \eqref{V1}. Physical variables are defined such that the difference between two neighboring physical variables is a physical fluctuation. Let two neighboring physical variables, $\hat g$ and $\hat g'=\hat g+\hat h$, differ by an infinitesimal $\hat h$. Physical variables $\hat g$ are defined such that $\hat h$ is a physical fluctuation, $\hat h=f$. 

Formally, physical variables obey the constraint
\be\label{V4}
V^T(\hat g)\hat g=0.
\ee
Here $V(g)$ obeys the differential relation
\be\label{123A}
V_i+\frac{\partial V_k}{\partial g_i}g_k=v_i~,~P_j{^i}v_i=0.
\ee
Applying the constraint on two neighboring physical variables yields for the difference
\be\label{132A}
V_i(\hat g+\hat h)(\hat g+\hat h)_i-V_i(\hat g)\hat g_i=0.
\ee
For infinitesimal $\hat h=h$ this implies
\be\label{132B}
\left(\frac{\partial V_i}{\partial \hat g_k}\hat g_i+V_k\right)h_k=v_kh_k=0.
\ee
Comparing with eq.\eqref{CB9}, $h=f+a$, and using, 
\ba\label{132C}
v_k(f_k+a_k)&=&v_k\big[(P h )_k+\big((1-P) h \big)_k\big]\nn\\
&=&v_k\big[(1-P) h \big]_k=v_ka_k,
\ea
eq. \eqref{132B} implies $a_k=0$ and we identify indeed
\be\label{132D}
\hat h=f.
\ee

The family of physical variables $\hat g$ corresponds to a ``trajectory'' in the space of variables $ g $ where two neighboring points are connected by a physical fluctuation $f$. Since the trajectory only has to obey a differential equation the family of physical variables is specified uniquely if a suitable initial value $\bar g_0$ is chosen. From there it can be spanned by subsequent additions of physical fluctuations $f$.

\subsection{Physical and gauge variables from trajectories in}

~{\bf field space}

For our simple example the physical variables $\hat g$ are represented by a line in the two-dimensional space of macroscopic fields $ g $. This line is specified by a choice of an initial condition $\bar g_0$. Two neighboring $\bar g_0$ are equivalent if they generate the same trajectory $\hat{g}$ for physical fields. This is realized if the two neighboring initial values $\bar{g}_0$ differ by a physical fluctuation. In contrast, different lines of physical metrics are induced if the two neighboring initial values differ by a gauge fluctuation. 

In the two-dimensional space spanned by $g=(g_1,g_2)$ the physical variables constitute a line that may be parametrized by some parameter $\sigma$, e.g. $\hat g(\sigma)$. This line is defined by a differential equation with initial condition $\hat g(\sigma=0)=\bar g_0$,
\be\label{ZV1}
\Big(1-P\big (\hat g(\sigma)\big)\Big)\partial_\sigma\hat g(\sigma)=0~,~\hat g(\sigma=0)=\bar g_0.
\ee
This differential equation ensures that an infinitesimal difference between two points on the line constitutes a physical fluctuation. 
We have depicted this line schematically in fig. 1. A different value for $\bar g_0$ will lead, in general, to a different line. 

We next define $\hat c$, again as a curve $\hat c(\sigma,\tau)$ solving a differential equation with initial value
\be\label{ZV2}
P\big(\hat g(\sigma)+\hat c(\sigma,\tau)\big)\partial_\tau\hat c(\sigma,\tau)=0~,~\hat c(\sigma,\tau=0)=0.
\ee
Thus for every $\sigma$ one can construct $\hat c(\sigma,\tau)$ by starting at $\tau=0$ with $\hat c=0$, adding an (infinitesimal) gauge fluctuation $a_1$ that obeys $P(\hat g)a_1=0$, subsequently adding a gauge fluctuation $a_2$ obeying $P(\hat g+a_1)a_2=0$, and so on. The line $\hat c(\sigma,\tau)$ exists for every $\sigma$. For every parameter combination $(\sigma,\tau)$ we can define $g(\sigma,\tau)=\hat g(\sigma)+\hat c(\sigma,\tau)$. If there are no bifurcation points arbitrary $g$ can be realized as $g(\sigma,\tau)$ for suitable $\sigma$ and $\tau$. We show the lines $g(\sigma,\tau)$ for different fixed $\sigma$ in fig. 1. Special values are $g(0,0)=\bar g_0,g(\sigma,0)=\hat g(\sigma)$. 

Inversely, for a given $g$ we can follow the differential equation \eqref{ZV2} for $\hat c$ until the line $g-\hat c(\tau)$ intersects the line of physical variables $\hat g(\sigma)$ for some $\sigma$. For this purpose $\sigma$ needs not to be known - the projector in eq. \eqref{ZV2} only depends on $g(\sigma,\tau)$, and for fixed $\sigma$ and $\hat g(\sigma)$ we have the differential equation $P\big(g(\sigma,\tau)\big)\partial_\tau g(\sigma,\tau)=0$. This procedure can be visualized in fig. 1. For a fixed $g$, indicated by the filled square, one follows the trajectory for fixed but unknown $\sigma$ until it intersects the line of physical metrics (open square), thus determining $\hat g(\sigma)$, and correspondingly $\hat c(\tau,\sigma)$. For every $g$ we can construct in this way $\hat g(g)$ and $\hat c(g)$ and realize the decomposition \eqref{V1}. We indicate the variables $\hat g(g)$ and $\hat c(g)$ by arrows in fig. 1. In the following we will assume the absence of bifurcations, at least in the region close to $\hat g(\sigma)$ which will be needed for our purposes. With this assumption the decomposition \eqref{V1} exists and is unique. 

In principle, the choice of the initial value $\bar{g}_0$, that is needed for a unique specification of $\hat{g}(g)$, is arbitrary. An obvious possible choice is to identify $\bar{g}_0$ with the expectation value $<g'>$ in the absence of sources,

\be\label{138A}
\bar{g}_0=\kl g'\kr_{|L=0}.
\ee
The difference between $\hat{g}$ and $<g'>$ arises then only for $L\neq0$.

\subsection{Differential constraints}

We can use this construction in order to derive two differential constraints for the map $g\to \hat g(g)$. An infinitesimal change $d\hat g=\hat h$ obeys $P(\hat g)d\hat g=d\hat g$, and we infer
\be\label{154A}
P(\hat g)\frac{\partial \hat g}{\partial g}=\frac{\partial \hat g}{\pg},
\ee
or
\be\label{97A}
P_i{^j}(\hat g)\frac{\partial\hat g_j}{\partial g_k}=\frac{\partial \hat g_i}{\partial g_k}.
\ee
An infinitesimal gauge variation of $g$, $dg=a$, obeys $P(g)dg=0$. It changes $\hat c$, corresponding to a change of $\tau$ at fixed $\sigma$, but leaves $\hat g$ invariant,
\be\label{154B}
d\hat g=\frac{\partial \hat g}{\pg}dg=\frac{\phg}{\pg}\big(1-P(g)\big)dg=0.
\ee
This entails the constraint
\be\label{154C}
\frac{\phg}{\pg}P(g)=\frac{\phg}{\pg},
\ee
or 
\be\label{99A}
\frac{\partial \hat g_k}{\partial g_i}P_i{^j}(g)=\frac{\partial \hat g_k}{\partial g_j}.
\ee
By construction, the physical variable $\hat g$ is gauge invariant. This obtains formally from eq. \eqref{CB5} by virtue of the constraint \eqref{154C},
\ba\label{154D}
\delta_\xi\hat g&=&\frac{\phg}{\pg}\delta_\xi g=-\xi\frac{\phg}{\pg}v(g)\nn\\
&=&-\xi\frac{\phg}{\pg}P(g)v(g)=0.
\ea

Similarly, a physical variation of $g,dg=f$, changes $\hat g$ and leaves $\hat c$ invariant. This implies the relation
\be\label{100A}
\frac{\partial \hat c}{\partial g}P(g)=0~,~\frac{\partial \hat c_k}{\partial g_i}P_i{^j}(g)=0.
\ee
We are interested in the region of very small $|\hat c|$, with typically values that vanish as $\alpha$ goes to zero. Infinitesimal $\hat c$ obey the simple condition
\be\label{ZV3}
P(\hat g)\hat c=0.
\ee
We can identify $\hat c$ with a gauge fluctuation around a ``background'' $\bar g=\hat g$. The constraint \eqref{ZV3} holds, however, only for infinitesimal $\hat c$, 
\be\label{155A}
\big(1-P(\hat g)\big)\hat c=\hat c+0(\hat c^2).
\ee

\subsection{Gauge invariant variable}

The physical variable $\hat g$ is a two-component variable. Due to the differential constraint \eqref{99A} it is gauge invariant. The valley corresponding to $\hat g(g)$ is a one-dimensional hypermanifold. We may therefore construct a single gauge invariant variable $s$ corresponding to this hypermanifold. 

For this purpose we employ normalized vectors $w$ and $v$ that are eigenstates of the projector $P$,
\ba\label{Z7}
P w=w~,~(1-P)v=v,
\ea
with 
\be\label{E9B}
w^T w=1~,~v^T v=1~, w^T v = 0.
\ee
For the projector \eqref{CB1} they are given explicitly by
\be\label{E9C}
w=\frac{1}{\sqrt{1+\eta^2}}\binom 1\eta~,~v=
\frac{1}{\sqrt{1+\eta^2}}\binom{-\eta}{~1}.
\ee
The eigenvectors of $P$ obey
\be\label{112A}
w_i w_j = P_{i j}, \quad v_i v_j = \left(1 -P \right)_{i j}.
\ee

We further introduce vectors $U$ and $V$ that are related to $w$ and $v$ by a differential relation
\be\label{E9E}
w_i = U_i + \frac{\partial U_k}{\partial g _i} g _k,\quad  v_i = V_i + \frac{\partial V_k}{\partial g _i} g _k.
\ee
The difference between $U$ and $w$ or $V$ and $v$ reflects the dependence of $w$ and $v$ on $ g $, which arises since $\eta$ depends on $g$. The gauge invariant variable $s$ is constructed as
\be\label{XY}
s=U_kg_k.
\ee
Indeed, $U^T(g)g$ is invariant,
\ba\label{GE6}
\delta_\xi (U_k(g)g_k)&=&\left(U_i+\frac{\partial U_k}{\partial g_i}g_k\right)\delta_\xi g_i\nn\\
=w_i\delta_\xi g_i&=&-\xi w_iv_i=0.
\ea
The gauge invariant variable $s$ can be extended to gauge invariant field combinations in Yang-Mills theories or gravity. This is mainly an argument of existence, since an explicit construction of $U$ may be difficult.

\begin{figure}[h!tb]
\begin{center}
\includegraphics[width=0.5\textwidth]{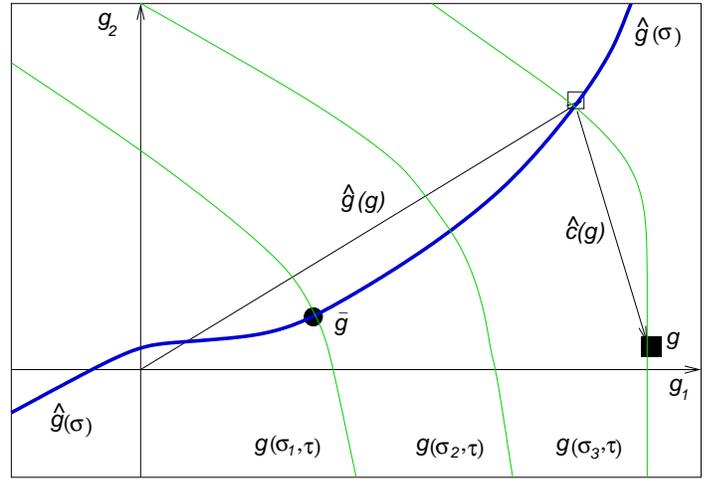}
\caption{Trajectories for physical variables $\hat g$ (thick line) and gauge variables $\hat c$ (thin lines). The function $\hat g(g)$ obtains by following for a given $g$ (filled square) the trajectory $g(\sigma,\tau)$ until it intersects the trajectory $\hat g(\sigma)$ (open square). We show the corresponding vectors $\hat g(g)$ and $\hat c(g)$.}
\label{giea_fig1}
\end{center}
\end{figure}

\section{Macroscopic gauge symmetry}
\label{Macroscopic gauge symmetry}

\subsection{Gauge invariance}

Let us consider an effective action $\Gamma(g)$ whose precise definition from microscopic physics (e.g. the relation with $\tilde \Gamma(g)$ and the specification of $L(g)$) is not important here. Using the decomposition \eqref{V1}, $g=\hat{g}+\hat{c}$, we first define
\be\label{137A}
\bar{\Gamma}(\hat{g})=\Gamma(\hat{g},\hat{c}=0).
\ee
This can be interpreted as the effective action for the light variable, with heavy variable $\hat c$ eliminated. At this point $\hat g$ is a constrained variable. 

The effective action for the light variable can be extended to the gauge invariant effective action 
\be\label{137B}
\bar{\Gamma}(g)=\bar{\Gamma}(\hat{g}(g)).
\ee
This description is redundant since $\G(g)$ depends formally on an arbitrary variable $g$ despite the fact that it only involves $\hat g$. Gauge symmetry expresses this redundancy. By virtue of the differential constraint \eqref{99A} one has 
\be\label{137C}
\frac{\partial\bar{\Gamma}}{\partial g_k}=\frac{\partial\bar{\Gamma}}{\partial\hat{g}_j} \frac{\partial\hat{g}_j}{\partial g_k} = \frac{\partial\bar{\Gamma}}{\partial\hat{g}_j} \frac{\partial\hat{g}_j}{\partial g_i} P_i^k=\frac{\partial\bar{\Gamma}}{\partial g_i} P_i^k.
\ee
The gauge variation of $\bar\Gamma$ therefore vanishes
\be\label{137D}
\delta_{\xi}\bar\Gamma=\frac{\partial\bar\Gamma}{\partial g_k} \delta_{\xi}g_k= -\xi \frac{\partial\bar\Gamma}{\partial g_k} v_k = - \xi\frac{\partial\bar\Gamma}{\partial g_i} P_i^k v_k=0.
\ee
Of course this is a direct consequence of the gauge invariance of the physical variable $\hat{g}$. From eq. \eqref{137C} one also concludes that the first variation of $\bar\Gamma$ is some generalized physical source

\be\label{137E}
\frac{\partial\bar\Gamma}{\partial g_k}=\bar{J}^k ,~\bar{J}^iP_i^k=0.
\ee

The definition \eqref{137A} is motivated if $\Gamma$ involves a large quadratic term
\be\label{166A}
\Gamma_{gf}=\frac{1}{2\alpha}\hat c_k\hat c_k,
\ee
For $\alpha\rightarrow0$ the field equations derived by variation of $\Gamma$ will be solved for $\hat{c} =0$. Setting $\hat{c}=0$ eliminates the heavy mode, such that $\bar\Gamma$ is indeed the effective action for the light mode.

We may use fig. 1 in order to visualize our construction of the gauge invariant effective action $\bar\Gamma(g)$. The effective action $\Gamma(g)$ (without gauge symmetry) is a function over the $(g_1,g_2)$ plane. For small $\alpha$ it has a deep valley along the line of the physical variable $\hat g$, with second derivative perpendicular to this line $\sim 1/\alpha$. The action $\bar \Gamma(\hat g)$ has support only on the line for the variable $\hat g$. The extension $\bar\Gamma(g)=\bar\Gamma\big(\hat g(g)\big)$ has again support in the whole plane. It is constant along the lines $g(\sigma,\tau)$ with fixed $\sigma$, taking the value $\bar\Gamma\big(\hat g(\sigma)\big)$.

\subsection{Macroscopic emergence of gauge symmetry}

At this point we may formulate the general condition how an effective gauge invariant theory arises from a more general effective action $\Gamma(g)$ that is not gauge invariant. Two conditions are sufficient: (i) $\Gamma(g)$ contains a term $\sim \hat c^2$ with a coefficient $\sim 1/\alpha$ that exceeds all other relevant scales. Here $\hat c$ is defined by the decomposition $g=\hat g+\hat c$, with $\hat g$ obeying a differential constraint. We will take later $\alpha\to 0$ such that small $\hat c$ obeys in linear order $P(\hat g)\hat c=0$. (ii) A possible term linear in $\hat c$ in $\Gamma(g)$ should have a coefficient that remains finite for $\alpha\to 0$. These two conditions are realized if $\Gamma(g)$ contains a gauge fixing term \eqref{166A}
and no other terms diverge $\sim\alpha^{-1}$. 

As a consequence of these two conditions the field equations can be projected into a ``heavy'' and a``light'' sector, 
\be\label{GE1}
\big(1-P(g)\big)\frac{\partial\Gamma}{\partial g}=\big(1-P(g)\big)\bar L=\bar H\sim \frac1\alpha\hat c+\dots 
\ee
and 
\be\label{GE2}
P(g)\frac{\partial\Gamma}{\partial g}=P(g)\bar L=\bar J=A_0(\hat g)+A_1(\hat g)\hat c+\dots, 
\ee
where $g=\hat g+\hat c$. Here we use the relation \eqref{100A}, and eq. \eqref{GE1} reads more precisely
\be\label{121a}
H_i=\frac1\alpha\hat c_k\frac{\partial\hat c_k}{\partial g_i}+\dots
\ee
The coefficients $A_0,A_1$ for the expansion in powers of $\hat c$ in eq. \eqref{GE2} remain finite for $\alpha\to 0$. The dots on the r.h.s. of eqs. \eqref{GE1}, \eqref{GE2} denote terms that vanish for $\hat c\to 0$ if $\hat c/\alpha$ is finite. For $\alpha\to 0$, and finite sources $L$, the solution of eq. \eqref{GE1} implies $\hat c=0$. This can be inserted into eq. \eqref{GE2} such that only $A_0(\hat g)$ matters. For the light degrees of freedom $\hat g$ we can define $\bar\Gamma(\hat g)=\Gamma(g,\hat c=0)$. The gauge invariant effective action $\G(g)$ obtains then by dropping the constraint on $\hat g$ and extending $\bar\Gamma(g)=\bar\Gamma\big (\hat g(g)\big)$. 

We next show that for $\hat c=0$ the l.h.s of eq. \eqref{GE2} can be written as $\partial \bar\Gamma/\partial g$, 
\be\label{GE3}
P(g)\frac{\partial\Gamma(g)}{\partial g}_{|\hat c=0}=\frac{\partial\bar\Gamma(g)}{\partial g}.
\ee
For this purpose we write
\be\label{122A}
\Gamma(g)=\G\big(\bar g(g)\big)+\hat c_kB_k,
\ee
such that 
\be\label{122B}
\frac{\partial \Gamma}{\partial g_j}_{|\hat c=0}=\frac{\partial\G}{\partial g_j}+\left(\frac{\partial \hat c_k}{\partial g_j}B_k\right)_{|\hat c=0}.
\ee
Eq. \eqref{GE3} follows from eqs. \eqref{137C} and \eqref{100A}. We recover eq. \eqref{137E}. 

We emphasize that the conditions leading to a gauge invariant effective action $\bar\Gamma(g)$ for the light degrees of freedom are purely formulated on the ``macroscopic level'', e.g. in terms of the effective action $\Gamma(g)$. No particular assumption on the microscopic physics that leads to these conditions is required. This opens the possibility that gauge invariance emerges as a result of the ``renormalization flow'' from microphysics to macrophysics. It is sufficient that this flow produces or keeps a term $\sim \hat c^2/\alpha$ that is huge on the scales of the effective theory for the light degrees of freedom, and that no huge term linear in $\hat c$ is generated. The precise relation between $\Gamma(g)$ and $\tilde \Gamma(g)$, as defined by eq. \eqref{85A} is not important. Also the sources $\bar L,\bar J, \bar H$ may differ from $L,J,H$.

\subsection{Gauge symmetry from microscopic formulation}

The crucial point for the emergence of macroscopic gauge symmetry from the decoupling of the heavy degree of freedom is the presence of the gauge fixing term \eqref{166A} in $\Gamma$, with $\alpha\to 0$. We will show that this obtains in the microscopic formulation \eqref{Z1}. The functions $\Gamma(g)$ and $\tilde \Gamma(g)$ may differ by a term that does not diverge for $\alpha\to 0$. This difference will not affect the divergent gauge fixing term \eqref{166A} in $\Gamma(g)$. 

We identify the dynamical background $\bar g(g)$ with $\hat g(g)$. For the leading divergent term for $\alpha\to 0$ the saddle point approximation becomes valid, such that 
\be\label{RA}
\tilde\Gamma(g)=\tilde \Gamma_{gf}+\tilde \Gamma_{fin},
\ee
where 
\be\label{RB}
\tilde\Gamma_{gf}=\frac{1}{2\alpha}\kl c'_k\kr \kl c'_k\kr,
\ee
and $\tilde \Gamma_{fin}$ does not diverge $\sim \alpha^{-1}$ for $\alpha\to 0$. Indeed, $\tilde \Gamma_{gf}$ corresponds to the ``classical approximation'', while ``loop-corrections'' do not diverge $\sim \alpha^{-1}$. The expectation value $\kl c'\kr (g)$ is evaluated in the presence of sources and obeys, by virtue of eq. \eqref{75B}, 
\be\label{RBB}
P(g)\kl c'\kr (g)=0.
\ee
The precise relation between $g$ and $L$ does not matter in this context.

For small $\hat c$ we can expand the propagator $P(g)=P(\hat g+\hat c)$ in $\hat c$. The leading term in the constraint \eqref{RBB} takes the form 
\be\label{RC}
P(\hat g)\kl c'\kr(g)=0,
\ee
which coincides with the projector equation \eqref{ZV3} for infinitesimal $\hat c$. This suggests that for a suitable choice of the macroscopic variable $g$ the quantities $\kl c'\kr(g)$ and $\hat c(g)$ can be identified in the limit $\alpha\to 0$. 

The relation $g(L)$ can be defined implicitly by the relation between $g$ and $\kl g'\kr$. We choose the macroscopic variable $g$ such that 
\be\label{RD}
g=\kl g'\kr-\kl b'\kr.
\ee
With this choice and $\bar g=\hat g$ the relations
\be\label{RE}
\kl g'\kr -\hat g=\kl b'\kr+\kl c'\kr~,~g-\hat g=\hat c,
\ee
indeed imply $\hat c=\kl c'\kr$. This closes the argument that $\Gamma(g)$ contains a gauge fixing term \eqref{166A} in the limit $\alpha\to 0$, and therefore establishes the gauge invariant effective action $\bar \Gamma(g)$ for the light degree of freedom. 

We observe that the condition \eqref{RD} does not fix the choice of the macroscopic variable $g$ uniquely. One possible choice could simply be $g=\kl g'\kr$, which is equivalent to $\kl b'\kr=0$. We will admit, however, the more general choice \eqref{RD}, for which we only require the condition
\be\label{RF}
P(g)(g-\kl g'\kr)=g-\kl g'\kr.
\ee
One may use the remaining freedom in the choice of $g(L)$ (compatible with eq. \eqref{RF}) and the precise relation between $\Gamma$ and $\tilde \Gamma$ in order to ``optimize'' the properties of $\bar \Gamma(g)$. For example, this freedom is used in ref. \cite{CWFLOW} in order to obtain a simple form of a gauge invariant flow equation for a scale-dependent $\G[g]$. 

We conclude that macroscopic gauge symmetry can emerge from microphysics in a rather general setting. The microscopic formulation \eqref{Z1} is only an example for a much wider class of microscopic settings that can lead to macroscopic gauge invariance. For $\alpha\to 0$ the precise form of the microscopic action $S$ is arbitrary. The only thing that fixes the gauge symmetry is the form of the diverging effective action $\Gamma_{gf}$ for the heavy degree of freedom. In the context of flow equations it is sufficient that $\alpha$ flows towards zero, even if it does not vanish on the microscopic level. The generic emergence of gauge symmetry for the effective action $\bar \Gamma(g)$ does not yet guarantee that the properties of $\G(g)$ are simple. For quantum field theories this concerns, in particular, locality properties of $\bar \Gamma(g)$.

\section{Multi-component variables}
\label{Multi-component variables and fields}

In this section we proceed towards the construction of the gauge invariant effective action for quantum field theories. The main conceptual issues can already be understood by the two-component examples of the preceding two sections. The way to quantum field theory proceeds by a rather straightforward generalization to $N$-components, and finally to the limit $N\to\infty$. The indices of the $N$-component vectors will then contain spacetime coordinates or momenta.

\subsection{Multi-component vectors}\label{multi-component}

It is straightforward to generalize the microscopic and macroscopic variables $g'$ and $g$, as well as the sources $L$, to $N$-component vectors.  The projectors $P$ and $(1-P)$ depend again on the macroscopic variable $g$. The number of eigenvalues $1$ and $0$ of $P$ needs not to be equal. If $P$ has $M$ eigenvalues $0$ we have $M$ ``gauge degrees of freedom", $c' = \left( 1 - P \right) h'$, and $N-M$ ``physical degrees of freedom", $b' = Ph'$. The projector is not necessarily symmetric. We use covariant vectors $g_i,g'_i$ and contravariant vectors $L^i$ for the sources, such that 
\ba\label{197}
-S_L&=&L^ig'_i=L^i\bar g_i+J^ib'_i+H^ic'_i\nn\\
&=&L^Tg'=L^T\bar g+J^Tb'+H^Tc'.
\ea
The physical sources $J^i$ obey 
\be\label{197A}
J^i=L^iP_i{^j},~J^iP_i{^j}=J^j,~J^TP=J^T.
\ee
As before, the relation
\be
\frac{\partial \bar{\Gamma}}{\partial g_i} = J^i
\ee
implies gauge invariance under infinitesimal transformations 
\be\label{Q3}
\delta g = \left( 1 - P \right) \lambda,~\delta g_i=(1-P)_i{^j}\lambda_j,
\ee
according to 
\be
\delta \bar{\Gamma}=J^i\delta g_i = J^T \delta g = J^T P \left( 1 - P \right) \lambda = 0.
\ee

The projector $1-P$ has $M$ eigenvectors $v^s$ for the eigenvalue one, 
\be
\left( 1 - P \right)_i^{\hspace{0,2cm} j} v_j^s = v_i^s,
\ee
and $N-M$ eigenvectors $w^u, u = 1 \cdots N-M$, for the eigenvalues zero (or eigenvalues one of $P$),
\be
P_i^{\hspace{0,2cm} j} w_j^u = w_i^u.
\ee
In terms of these eigenvectors the gauge transformation takes the form 
\be\label{204A}
\lambda_i=-v_i{^s}\xi_s,~\delta g_i=-v^s_i\xi_s.
\ee

Contravariant vectors are related to covariant vectors by 
\be
g'^i = D^{i j} g'_j, ~L^i = L_j D^{j i}
\ee

\noindent
(For symmetric projectors $P^T = P$ one can use $D^{i j} = \delta^{i j}$ such that there is no difference between $g'^i$ and $g'_i$.)
We also employ 

\begin{align}
v^{s,i} = D^{i j} v_j^s, \quad v_s^i = D^{i j}v_j^t F_{t s}
\end{align}

\noindent
and similarly for $w_u^i$, obeying 

\begin{align}
v_s^i \left( 1 - P \right)_i^{\hspace{0,2cm} j} = v_s^j, \quad w_u^i P_i^{\hspace{0,2cm} j} = w_u^j.
\end{align}

\noindent
(For $P^T = P$ we can choose $ F_{t s} = \delta_{t s}$.) The normalization is chosen as

\begin{align}\label{Q8}
v_t^i v_i^s = \delta_t^s, \quad w_u^i w_i^v = \delta_u^v,
\end{align}

\noindent
and the orthogonality of eigenspaces implies 

\begin{align}
v_i^s w_u^i = 0.
\end{align}

\noindent
Using $v_s^i v_i^s = M$, $w_u^i w_i^u = N-M $, as well as the projector properties, one finds
\be
v_i^s v_s^j = \left( 1 - P \right)_i^{\hspace{0,2cm} j}, \quad w_i^u w_u^j = P_i^{\hspace{0,2cm} j}.
\ee

\subsection{Physical and gauge degrees of freedom}

We will employ again the physical and gauge degrees of freedom $\hat g_i$ and $\hat c_i$ and the decomposition
\be\label{PG5}
g_i=\hat g_i+\hat c_i.
\ee
Their construction is analogous to the two-variable model of sect. \ref{Physical variable}. For any given $g_i$ the physical and gauge fluctuations $f_i$ and $a_i$ obey 
\be\label{PG6}
\big(1-P(g)\big)_i{^j}f_j=0~,~P(g)_i{^j}a_j=0.
\ee
Starting from a given initial value $\bar g_{0,i}$ we subsequently add physical fluctuations $f_i$ in order to construct the $N-M$-dimensional hypermanifold $\hat g_i$. This manifold may be parametrized by $N-M$ variables $\sigma_u$ as $\hat g_i(\sigma_u)$. The physical degrees of freedom obey the differential equation
\be\label{PG7}
\big (1-P(\hat g)\big)_i{^j}\frac{\partial\hat g_j}{\partial\sigma_u}=0~,~\hat g_j(\sigma_u=0)=\bar g_j.
\ee
For every $\hat g_i(\sigma_u)$ we then construct $\hat c_i(\sigma_u,\tau_s)$ by subsequently adding gauge fluctuations. They are solutions of the differential equations
\be\label{PG8}
P_i{^j}(\hat g+\hat c)\frac{\partial\hat c_j}{\partial \tau_s}=0~,~\hat c_j(\sigma_u,\tau_s=0)=0.
\ee
We again assume the absence of bifurcations in the relevant region of small $\hat c$ such that the decomposition \eqref{PG5} exists and $\hat g(g)$ and $\hat c(g)$ are unique. 

The map $g\to\hat g$ obeys simple differential properties. First, an infinitesimal difference between two physical degrees of freedom is a physical fluctuation. It therefore obeys
\be\label{PG8A}
d\hat g_i=P_i{^j}(\hat g)d\hat g_j=P_i{^j}(\hat g)
\frac{\partial \hat g_j}{\partial g_k}d g_k,
\ee
implying for the derivatives
\be\label{PG8B}
\frac{\partial\hat g_i}{\partial g_k}=P_i{^j}(\hat g)
\frac{\partial \hat g_j}{\partial g_k}.
\ee
A simple constraint
\be\label{PG8C}
\big(\delta^j_i -P_i{^j}(\hat g)\big)
\frac{\partial \hat g_j}{\partial g_k}=0
\ee
therefore applies to the partial derivatives $\partial\hat g_j/\partial g_k$, and not to $\hat g_j$. 

Second, changing $g_j$ by a gauge fluctuation changes $\hat c_i$ but does not affect $\hat g_i$,
\be\label{PG8D}
\hat g_i\big(g_j+(1-P)_j{^k}dg_k)=\hat g_i(g_j).
\ee
This implies the constraint
\be\label{PG8E}
\frac{\partial\hat g_i}{\partial g_k}
\big(\delta^j_k-P_k{^j}(g)\big)=0.
\ee
The differential relations \eqref{PG8C} and \eqref{PG8E} constrain the variation of the hyperface $\hat g(g)$ with $g$. They correspond to eq. \eqref{7Aa}. We also generalize the relation \eqref{100A}, such that infinitesimal $\hat c$ obey
\be\label{PG13}
P_i{^j}(\hat g)\hat c_j=0.
\ee

The infinitesimal gauge transformation of $ g_i$,
\be\label{PG10}
\delta_\xi g_i=-\xi_s v^s_i(g),
\ee
obeys
\be\label{PG11}
P_i{^j}\big(g)\big)\delta_\xi g_j=0.
\ee
This is a transformation in the gauge direction that can be realized by 
\be\label{PG12}
\delta_\xi\hat c_i=-\xi_s v^s_i(g)~,~\delta_\xi\hat g_i=0.
\ee
Indeed, the gauge invariance of $\hat g$ follows from the differential constraint \eqref{PG8E},(??),
\ba\label{A12A5}
\delta_\xi\hat g_i&=&\frac{\partial\hat g_i}{\partial g_j}\delta_\xi g_j=\frac{\partial\hat g_i}{\partial g_k}
P_k{^j}(g)\delta_\xi g_j\nn\\
&=&-\xi_s\frac{\partial\hat g_i}{\partial g_k}P_k{^j}(g)v^s_j(g)=0.
\ea
For infinitesimal $\hat c$ we can define
\be\label{PG14}
c_s=v^i_s(\hat g)\hat c_i.
\ee
These variables correspond to $c$ in the two-variable model. With the transformation \eqref{PG12} one has
\be\label{PG15}
\delta_\xi c_s=-\xi_s.
\ee

In conclusion, we have decomposed the macroscopic variables $g_i$ into gauge invariant physical variables $\hat g_i$ and gauge degrees of freedom $\hat c_i$. Gauge transformations only act on $\hat c_i$. The decomposition is, in general, not global and $\hat g_i$ only obeys differential constraints for its dependence on $g$. The precise definition $\hat g$ therefore depends on the choice of an ``initial value'' $\bar g_0$. 

We can again construct $N-M$ gauge invariant variables $s_u(g)$ which obey the defining relation
\be\label{PG1}
\frac{\partial s_u}{\partial g_i}=w^i_u(g),
\ee
and are formally given by
\be\label{PG2}
s_u=U^i_u(g)g_i,
\ee
with $U^i_u$ obeying the differential equation 
\be\label{PG3}
U^i_u+\frac{\partial U^k_u}{\partial g_i}g_k=w^i_u.
\ee
Both $s_u$ and $U^i_u$ are uniquely specified once the ``initial values'' for the solution of eq. \eqref{PG3} are given for some $\bar g_0$. The gauge invariance of $s_u$ follows from 
\ba\label{PG4}
\delta_\xi s_u&=&\frac{\partial s_u}{\partial g_i}\delta_\xi g_i=w^i_u\delta_\xi g_i\nn\\
&=&w^i_u(1-P)_i{^j}\lambda_j=0. 
\ea

Generalizing $g_i$ to fields the variables $s_u$ become gauge invariant field combinations. Their number corresponds to the physical degrees of freedom, as obtained by the number of degrees of freedom in $g$ minus the number of gauge degrees of freedom. While these gauge invariant field combinations exist, they are difficult to construct explicitly in practice.

The $M$-dimensional hypersurface spanned by $\hat g_i(\sigma_u)+\hat c_i(\sigma_u,\tau_s)$ for fixed $\sigma_u$ and $\hat g(\sigma_u)$ constitutes a manifold of constant $s_u(g)=U^i_u(g) g_i$, since 
\be\label{PG9}
ds_u=\frac{\partial s_u}{\partial g_i}h_i=w^u_ih_i
\ee
vanishes if $h_i$ is a gauge fluctuation. 

\subsection{Gauge invariant effective action}

Consider now an effective action $\Gamma(g)$ that contains a term 
\be\label{PG16}
\Gamma_{gf}=\frac{1}{2\alpha}\hat c_i T^{ij}\hat c_j.
\ee
We will take $\alpha\to 0$ and assume that no other parts in $\Gamma$ diverge in this limit. For $T^{ij}$ we assume that it has no zero eigenvalues on the projected space corresponding to $\hat c_i$ obeying eq. \eqref{PG13}. In other words, $\hat c_i T^{ij}\hat c_j=0$ implies $\hat c_i=0$. This is sufficient for the extraction of a gauge invariant effective action $\bar\Gamma(g)$. 

For finite sources the field equations
\be\label{PG17}
\frac{\partial\Gamma}{\partial g_i}=\bar L^i
\ee
require for the solution 
\be\label{PG18}
\hat c_i=0.
\ee
Inserting this partial solution into $\Gamma(g)$ yields the effective action for the light degrees of freedom $\hat g_i$, 
\be\label{PG19}
\bar\Gamma(\hat g_i)=\Gamma( g_i,\hat c_i=0).
\ee
The gauge invariant effective action $\bar\Gamma(g)$ is defined as the extension
\be\label{PG20}
\bar\Gamma(g)=\bar\Gamma\big(\hat g(g)\big).
\ee

We can use eq. \eqref{PG8E},
\be\label{PG21}
\frac{\partial\hat g_i}{\partial g_j}=\frac{\partial \hat g_i}{\partial g_k}P_k{^j}(g),
\ee
for establishing the gauge invariance of $\bar\Gamma(g)$. The first derivative obeys
\ba\label{PG22}
\frac{\partial\bar\Gamma}{\partial g_j}&=&\frac{\partial\G}{\partial \hat g_i}
\frac{\partial\hat g_i}{\partial g_j}=\frac{\partial\G}{\partial\hat g_i}
\frac{\partial\hat g_i}{\partial g_k}P_k{^j}(g)\nn\\
&=&\frac{\partial\G}{\partial g_k}P_k{^j}(g),
\ea
implying a vanishing gauge variation
\be\label{PG22a}
\delta_\xi\G=\frac{\partial \G}{\partial g_j}\delta_\xi g_j=-\xi_s
\frac{\partial\G}{\partial g_k}P_k{^j}(g)v^s_j(g)=0.
\ee

We conclude that a general effective action of the form 
\be\label{261A}
\Gamma(g)=\bar\Gamma(\hat g)+\frac{1}{2\alpha}\hat c_iT^{ij}\hat c_j+\Delta\Gamma(\hat g,\hat c)
\ee
is projected onto a gauge invariant effective action $\bar\Gamma(g)=\G\big(\hat g(g)\big)$ for the light degrees of freedom provided the limit $\alpha\to 0$ is taken. Here $\Delta\Gamma$ is assumed to remain finite for $\alpha\to 0$ (or diverge less fast than $\alpha^{-1}$), and it is defined such that it vanishes for $\hat c=0$. This projection is realized by the solution \eqref{PG18} of the field equation for the heavy degree of freedom. It corresponds to the ``decoupling of the heavy modes''. 

The ``functional'' integral
\be\label{261B}
Z(L)=\int \D g'\exp \left\{-S(g')-\frac{1}{2\alpha}c'_iT^{ij}(g)c'_j
+L^ig'_i\right\}
\ee
leads for $\alpha\to 0$ precisely to an effective action of the form \eqref{261A}. Gauge invariance of $S(g')$ is not needed. The gauge degrees of freedom are given by 
\be\label{261C}
\hat c_i=\kl c'_i\kr,
\ee
and the physical variables $\hat g$ are gauge invariant. They obey the differential constraints \eqref{PG8B}, \eqref{PG8E}. The argument proceeds in parallel to sect. \ref{Macroscopic gauge symmetry}, relying on the validity of the saddle point approximation for the leading singular term in the limit $\alpha\to 0$. 

We conclude that the construction of a gauge invariant effective action $\bar \Gamma(g)$ can be extended to an arbitrary number of fields. The field dependent projector $P(g)$ needs not to be symmetric. On the level of the ``functional'' integral the crucial ingredient is the form of the ``gauge fixing term'' \eqref{PG16}, being quadratic in the projected fluctuation fields $c'_i$, and therefore dependent through the projector on the macroscopic field $g$. The limit $\alpha\to 0$ leads to an effective action for the light fields. Its arguments are the physical fields $\hat g_i$, and $\G[\hat g]$ turns to a gauge invariant action if the constraint on $\hat g$ is dropped. The limit $N\to\infty $ does not pose any particular problem in this construction. We can therefore promote our construction to quantum field theories, and the integrals to functional integrals. 

\section{Quantum field theories}
\label{Quantum field theories}

The extension to quantum field theories is conceptually straightforward. It corresponds to the limit $N\to \infty$, where the index $i$ comprises now a space-time label $x^\nu$ as well as Lorentz and internal indices. 

\subsection{Yang-Mills theories}

For Yang-Mills theories the multi-component vector $g_i$ stands for the gauge field $A^z_\mu(x)$. The projector on physical modes $P(g)$ is given by \cite{RW}
\be\label{GA}
P_\mu{^\nu}(x,y)=\delta(x-y)(\delta^\nu_\mu-D_\mu D^{-2}D^\nu).
\ee
It depends on the macroscopic field $A_\mu$ through the covariant derivative
\be\label{GB}
D_\mu=\partial_\mu-iA_\mu.
\ee
Eqs. \eqref{GA} and \eqref{GB} involve matrices in the adjoint representation, e.g. 
\be\label{GC}
(A_\mu)_{yz}=-iA^w_\mu f_{wyz}.
\ee
For the particular case of abelian gauge theories one has $D_\mu=\partial_\mu$ and $P$ becomes independent of $A_\mu$. This generalizes the simple setting of sect. \ref{Gauge symmetry from decoupling}. 

For the decomposition
\be\label{GD}
A_\mu=\hat A_\mu+\hat c_\mu
\ee
the physical gauge fields $\hat A_\mu(A)$ obey the differential constraint 
\be\label{GE}
P_\mu{^\nu}(\hat A)\frac{\partial \hat A_\nu}{\partial A_\rho}=\frac{\partial \hat A_\mu}{\partial A_\nu}P_\nu{^\rho}
=\frac{\partial \hat A_\mu}{\partial A_\rho}.
\ee
(For non-abelian gauge theories no global relation of the type $P_\mu{^\nu}(\hat A)\hat A_\nu=\hat A_\mu$ is obeyed.) The gauge degrees of freedom $\hat c_\mu$ obey
\be\label{GF}
\frac{\partial \hat c_\mu}{\partial A_\nu}P_\nu{^\rho}(A)=0,
\ee
and for infinitesimal $\hat c_\mu$ one has
\be\label{GG}
P_\mu{^\nu}(\hat A)\hat c_\nu=0.
\ee

Assume now that $\Gamma[A]$ contains a gauge fixing term
\be\label{GH}
\Gamma_{gf}=\frac{1}{2\alpha}\int_xG^zG^*_z,~G^z=(D^\mu\hat c_\mu)^z.
\ee
For $\alpha\to 0$ the solution of the field equation
\be\label{GI}
\frac{\partial \Gamma}{\partial A^z_\mu(x)}=L^\mu_z(x)
\ee
is found for finite sources $L$ as
\be\label{GJ}
\hat c^z_\mu(x)=0.
\ee
From the effective action for the light modes
\be\label{GK}
\G[\hat A]=\Gamma[A,\hat c=0]
\ee
the gauge invariant action follows as
\be\label{GL}
\G[A]=\G[\hat A(A)].
\ee
By virtue of the differential constraints \eqref{GE} one has
\be\label{GM}
\frac{\partial\G}{\partial A_\mu}=J^\mu~,~P^\mu{_\nu}(A)J^\nu=J^\mu.
\ee
This is obeyed for covariantly conserved sources
\be\label{GN}
D_\mu J^\mu=0.
\ee
For a gauge variation
\be\label{GO}
\delta A_\mu=D_\mu\alpha,~P_\mu{^\nu}\delta A_\nu=0,
\ee
the conservation of the current \eqref{GN} implies gauge invariance of $\G$
\be\label{GP}
\delta\G\sim \int_x \tr \{J^\mu D_\mu\alpha\}=-\int_x\tr \big\{(D_\mu J^\mu)\alpha\big \}=0.
\ee

Starting from a microscopic formulation the macroscopic gauge fixing term \eqref{GH} is realized for a microscopic gauge fixing 
\be\label{GQ}
S_{gf}=\frac{1}{2\alpha}\int_x G'^zG'^*_z,
\ee
with 
\be\label{GR}
G'^z=\big[D^\mu(A)(A'_\mu-\hat A_\mu)\big]^z.
\ee
This corresponds to Landau gauge fixing with a dynamical background field. With 
\ba\label{GS}
A'_\mu&=&\hat A_\mu+b'_\mu+c'_\mu,~D^\mu b'_\mu=0,\nn\\
c'_\mu&=&D_\mu c',~c'=D^{-2}D^\nu(A'_\nu-\hat A_\nu),\nn\\
c'_\mu&=&(\delta^\nu_\mu-P_\mu{^\nu})(A'_\nu-\hat A_\nu),
\ea
one has
\be\label{GT}
D^\mu(A'_\mu-\hat A_\mu)=D^\mu c'_\mu,
\ee
and therefore
\be\label{GU}
G'^z=\big(D^\mu(A)c'_\mu\big)^z.
\ee
For $\alpha\to 0$ the leading order saddle point approximation generates in $\Gamma$ the required term \eqref{GH}, with 
\be\label{GV}
\hat c_\mu=\kl c'_\mu\kr.
\ee
We conclude that the macroscopic emergence of gauge symmetry is realized if the microscopic gauge fixing term takes the specific form of the Landau gauge fixing \eqref{GQ}, \eqref{GR}.

\subsection{Gravity}

For gravity the vector $g_i$ corresponds to the metric $g{\mn}(x)$. The explicit construction of the projector $P{\mn}{^{\rho\tau}}(x,y)$ on the physical fluctuations is more involved than for Yang-Mills theories. It has been discussed in ref. \cite{CWCFG}. We present here only the structural aspects.

The physical sources are denoted here by $K^{\mu \nu} = K^{\nu \mu}$ and obey the constraint, 
\be\label{K2}
\partial_{\mu} K^{\mu \nu} + \Gamma_{\mu \rho}^{\quad \nu}\left( g \right) K^{\mu \rho} = 0,
\ee
where $\Gamma_{\mu \rho}^{\quad \nu}\left(g \right)$ is the connection formed with the metric $g_{\mu \nu}$.

The effective action $\G$ will be constructed such that it obeys
\be\label{K1}
\frac{\partial \bar{\Gamma}}{g_{\mu \nu}} = K^{\mu \nu}.
\ee
With the constraint \eqref{K2} this results in diffeomorphism invariance of $\G$. 
Multiplying eq.($\ref{K1}$) with $\xi_{\mu}$ and performing an integration over $x$ yields after partial integration 

\begin{align}\label{K7}
\begin{split}
&\int_{x} \xi_{\mu} \left( \partial_{\nu} K^{\mu \nu} + \Gamma_{\nu \rho}^{\quad \mu} K^{\nu \rho} \right) \\
&=- \int_{x} K^{\nu \rho} \left( \partial_{\nu} \xi_{\rho} -\Gamma_{\nu \rho}^{\quad \mu} \xi_{\mu}\right) \\
&=- \int_x K^{\nu \rho} D_{\nu} \xi_{\rho} = 0.
\end{split}
\end{align} 

\noindent
Employing eq.($\ref{K1}$) and using the symmetry of $g_{\mu \nu}$ and $K^{\mu \nu}$ one obtains 

\begin{align}\label{K8}
\int_{x} \frac{\partial \bar{\Gamma}}{\partial g_{\mu \nu}} \delta_{\xi} g_{\mu \nu} = 0,
\end{align}

\noindent
where

\begin{align}\label{K9}
\begin{split}
\delta_{\xi} g_{\mu \nu} &= - \left( D_{\mu} \xi_{\nu} + D_{\nu} \xi_{\mu} \right) \\ &= - \partial_{\mu} \xi^{\rho} g_{\rho \nu} - \partial_{\nu} \xi^{\rho} g_{\mu \rho} - \xi^{\rho} \partial_{\rho} g_{\mu \nu}.
\end{split}
\end{align}

\noindent
With infinitesimal $\xi^{\mu}= g^{\mu \nu} \xi_{\nu}$ we recognize in eq.($\ref{K9})$ the variation of the metric with respect to an infinitesimal diffeomorphism transformation. 


The interpretation of the constraint (\ref{K2}) is straightforward. The energy momentum tensor $T^{\mu \nu}$ is related to $K^{\mu \nu}$ by 

\begin{align}\label{K10}
T^{\mu \nu} = \frac{2}{\sqrt{g}}K^{\mu \nu},
\end{align}

\noindent
such that the constraint (\ref{K2}) expresses the covariant conservation of the energy momentum tensor

\begin{align}\label{K11}
D_{\mu} T^{\mu \nu} = 0.
\end{align}

\noindent
A restriction to sources corresponding to conserved energy momentum tensors implies gauge invariance of the effective action. The latter can also be expressed by the local identity

\begin{align}\label{K12}
\partial_{\nu} \frac{\partial \bar{\Gamma}}{\partial g_{\mu \nu}} + \Gamma_{\nu \rho}^{\quad \mu} \frac{\partial \bar{\Gamma}}{\partial g_{\nu \rho}} = 0,
\end{align}

\noindent
which follows from the combination of eqs.(\ref{K1}) and (\ref{K2}). 
Inversely, local gauge invariance of $\bar{\Gamma}$ implies a conserved energy momentum tensor as reflected by the relation (\ref{K2}).


Let us next specify the projector $P$ on the physical fluctuations and sources. Consider for a given metric $\bar g_{\mu \nu}$ a neighbouring metric $\bar g_{\mu \nu} + h_{\mu \nu}$. We split the metric fluctuations $h_{\mu \nu}$ into ``physical fluctuations" $f_{\mu \nu}$ and ``gauge fluctuations" $a_{\mu \nu}$ ,

\begin{align}\label{K14}
h_{\mu \nu} = f_{\mu \nu} + a_{\mu \nu},
\end{align} 

\noindent
according to the decomposition into a vector and divergence free tensor part 

\be\label{K15}
a_{\mu \nu} = D_{\mu} a_{\nu} + D_{\nu} a_{\mu}, ~
D^{\mu}f_{\mu \nu} = 0.
\ee

\noindent
Here the covariant derivative $D_{\mu}$ involves the connection formed with $\bar g_{\mu \nu}$.
Also the lowering and raising of indices is performed with $\bar g_{\mu \nu}$ and $\bar g^{\mu \nu}$, respectively. An infinitesimal gauge transformation (\ref{K9}) of $\bar g_{\mu \nu}$ can be realized by $a_{\mu} \rightarrow a_{\mu} - \xi_{\mu}$, with invariant $f_{\mu \nu}$. This motivates the naming of the fluctuations $f_{\mu \nu}$ and $a_{\mu \nu}$. 

We can write the decomposition of $h$ formally in terms of the projector $P$, $P^2 = P$, namely 

\begin{align}\label{K16}
P a = 0, \quad P f = f.
\end{align}

\noindent
More explicitly, the projector is defined by two conditions: the first states that for arbitrary vectors $a_{\mu}$ one has 

\begin{align}\label{K17}
\int_{y} P_{\mu \nu}^{\quad \rho \tau}\left( x, y \right) \left( D_{\rho}a_{\tau} + D_{\tau}a_{\rho} \right) = 0,
\end{align}

\noindent
while the second expresses the projector property,

\begin{align}\label{K18}
\int_{y} P_{\mu \nu}^{\quad \rho \tau} \left( x, y \right) P_{\rho \tau}^{\quad \lambda \sigma} \left( y, z \right) = P_{\mu \nu}^{\quad \lambda \sigma} \left( x, z \right).
\end{align}

\noindent
Furthermore, $P$ obeys 

\begin{align}\label{K19}
P_{\mu \nu}^{\quad \rho \tau} = P_{\nu \mu}^{\quad \rho \tau} = P_{\mu \nu}^{\quad \tau \rho} = P_{\nu \mu}^{\quad \tau \rho}.
\end{align}

\noindent
With the definitions (\ref{K14}), (\ref{K15}) we can define the physical metric fluctuations by a projection from $h_{\mu \nu}$

\begin{align}\label{K20}
f_{\mu \nu}(x) = \int_{y} P_{\mu \nu}^{\quad \rho \tau}\left( x, y \right) h_{\rho \tau}\left( y \right).
\end{align}

\noindent
(This also relates $a$ to $h$ by $a = h - f = \left( 1 - P \right) h$.) The explicit construction of the projector is not as simple as for Yang-Mills theories, for a discussion see ref. \cite{CWCFG}.

We decompose the macroscopic metric $ g _{\mu \nu}$ into the ``physical metric'' $\hat g_{\mu\nu}$ and the gauge mode $\hat c_{\mu\nu}$

\begin{align}\label{K25}
 g _{\mu \nu} = \hat{g}_{\mu \nu} + \hat{c}_{\mu \nu}.
\end{align}  

\noindent
The family of physical metrics is characterized by the property that two neighbouring physical metrics differ by a physical fluctuation $f_{\mu \nu}$. This is a differential relation in function space
\be\label{229A}
P_{\mu\nu}{^{\rho\tau}}(\hat g)\frac{\partial \hat g_{\rho\tau}}{\partial g_{\alpha\beta}}
=\frac{\partial\hat g\mn}{\partial g_{\rho\tau}}P_{\rho\tau}{^{\alpha\beta}}(g)=\frac{\partial \hat g\mn}{\partial g_{\alpha\beta}}.
\ee
A unique manifold of solutions to the differential relations \eqref{229A} needs the specification of some ``initial value'' $\bar{g}^0_{\mu \nu}$, from which the other physical metrics can be obtained by subsequently adding physical fluctuations $f_{\mu \nu}$. Dealing with such a constraint in practice is rather cumbersome. Dropping the formal constraint will lead gauge symmetry and result in important practical simplification. 

It is important that the constraint on physical metrics is only formulated for the infinitesimal difference between two such metrics, $f_{\mu \nu ;}^{\hspace{0,3cm} \nu} = 0$. There is no corresponding ``global constraint" on $\hat{g}_{\mu \nu}$. For this reason the concept of a family of physical metrics is somewhat hidden A second reason is the differential character of the constraint which requires the choice of some initial $\bar{g}^0_{\mu \nu}$. Different $\bar{g}^0_{\mu \nu}$ lead to different families of physical metrics. The precise choice is arbitrary.

Using the decomposition \eqref{K25} we write an arbitrary functional of the macroscopic metric in the form
\be\label{32A}
\Gamma[g_{\mu\nu}]=\Gamma[\hat g{\mn},\hat c{\mn}].
\ee
Our setting is realized if $\Gamma \left[  g  \right]$ takes the form 

\begin{align}\label{K26}
\Gamma \left[  g _{\mu \nu} \right] = \bar{\Gamma} \left[\hat{g}_{\mu \nu} \right] + \Gamma_{g f} \left[ \hat{g}_{\mu \nu}, \hat{c}_{\mu \nu} \right],
\end{align}

\noindent
with a generalized ``gauge fixing term" $\Gamma_{g f}$ at least quadratic in $\hat{c}_{\mu \nu}$. The first derivative of $\Gamma$ yields sources $\bar L$,
\begin{align}\label{34A}
\frac{\partial \Gamma}{\partial  g _{\mu \nu}} =\bar L^{\mu \nu}.
\end{align}
The solutions with $\bar L^{\mu \nu} = K^{\mu \nu}$ correspond to $\hat{c}_{\mu \nu} = 0$, such that $\Gamma_{g f}$ vanishes if the solution is inserted.   This  extends to arbitrary $\bar L^{\mu\nu}$ if the coefficient $\sim \alpha^{-1}$ in front of the gauge fixing term diverges. As usual, the effective action for the ``light modes'' $\hat g_{\mu\nu}$ obtains from $\Gamma$ by inserting the solution of the field equation for the ``heavy modes'' $\hat c_{\mu\nu}$
\be\label{34Aa}
\bar\Gamma[\hat g{\mn}]=\bar\Gamma[\hat g{\mn},\hat c{\mn}=0].
\ee
The gauge invariant effective action $\bar{\Gamma}$ will then obtain by dropping the formal constraint on $\hat{g}_{\mu \nu}$,

\be\label{K27}
\bar{\Gamma} \left[ g_{\rho\sigma} \right] =\Gamma\big[\hat g{\mn}[g_{\rho \sigma}]\big].
\ee

\noindent
The gauge fixing term is no longer present.

We finally connect our setting to a gauge fixed functional integral with a particular gauge fixing 
\be\label{235A}
S_{gf}=\frac{1}{2\alpha}\int_x\sqrt{g}(D^\mu h'_{\mu\nu})^2,~h'_{\mu\nu}=g'_{\mu\nu}-\hat g_{\mu\nu}.
\ee
Here the covariant derivative $D_\mu$ is formed with the macroscopic metric $g_{\mu\nu}$. Together with its inverse $g^{\mu\nu}$ the macroscopic metric is also used to lower and raise indices. Decomposing
\ba\label{235B}
h'_{\mu\nu}&=&b'_{\mu\nu}+c'_{\mu\nu},\nn\\
b'_{\mu\nu}&=&P_{\mu\nu}{^{\rho\tau}}(g)h'_{\mu\nu},~D^\mu b'_{\mu\nu}=0,\nn\\
c'_{\mu\nu}&=&D_\mu c'_\nu+D_\nu c'_\mu,
\ea
one finds that $S_{gf}$ is indeed quadratic in $c'\mn$ and does not involve $b'\mn$,
\be\label{235C}
S_{gf}=\frac{1}{2\alpha}\int_x\sqrt{g}(D^\mu c'\mn)^2.
\ee
Identifying 
\be\label{ID}
\hat c\mn=\kl c'\mn\kr
\ee
induces in the effective action the gauge fixing term $\Gamma_{gf}$, for which $c'$ is replaced by $\hat c$ in eq. \eqref{235C}. The particular ``physical gauge fixing'' \eqref{235A} has been advocated in ref. \cite{CWCFG} and used for flow equations in ref. \cite{Gi}.

\section{Conclusions}
\label{Conclusions}

We have investigated the possibility that local gauge symmetries emerge macroscopically from microscopic laws that do not necessarily exhibit these symmetries. As a basic concept, the flow of a scale dependent effective action from short distances (microphysics) to large distances (macrophysics) may generate the gauge symmetries. This seems indeed possible. 

The mechanism for the dynamical generation of a local symmetry differs, however, profoundly from the case of a global symmetry. A general effective action $\Gamma$ may be written as a gauge invariant part $\G$ and a gauge violating part $\Delta \Gamma$,
\be\label{S1}
\Gamma(g)=\G(g)+\Delta\Gamma(g).
\ee
For a dynamical emergence of a global symmetry $\Delta\Gamma$ should flow towards zero in the infrared. In contrast, a local symmetry can be realized dynamically if $\Delta\Gamma$ diverges in the infrared in a particular way. This happens if $\Delta\Gamma$ constitutes a gauge fixing term which is quadratic in the gauge fluctuations. The quadratic term separates heavy from light modes. Eliminating the heavy modes eliminates $\Delta\Gamma$, leaving $\G$ as the effective action for the light modes. At this point the light modes correspond to constrained fields $\hat g(g)$. Dropping the constraint by the extension $\G(g)=\G\big(\hat g(g)\big)$ results in a redundant description that exhibits the gauge symmetry. 

This mechanism works for a particular form of a diverging gauge fixing term $\Delta\Gamma\sim \Gamma_{gf}$. The gauge fixing term defines a hypermanifold of light fields $\hat g(g)$ for which it does not contribute. The gauge modes are perpendicular to this hypermanifold, and the gauge invariance of $\G(g)$ expresses that $\G$ does not depend on the gauge modes. The characterization of the gauge modes, and the corresponding projector $P$ on physical fluctuations, is determined by the form of $\Gamma_{gf}$. For a given gauge symmetry, as a local gauge group for Yang-Mills theories or diffeomorphisms for gravity, the properties of physical fields $\hat g(g)$ and gauge modes $\hat c(g)$ are fixed (up to some initial values). This corresponds to a particular class of ``physical'' gauge fixing terms that produce for the light fields the wanted gauge symmetry. For Yang-Mills theories the Landau gauge with dynamical background field belongs to this class. For gravity the covariant conservation $D^\nu h{\mn}=0$ for metric fluctuations $h{\mn}$ is a physical gauge fixing. Other gauge fixing terms may lead to a projection on heavy and light modes that do not correspond to the wanted gauge symmetry. 

It is possible that the required gauge fixing term is generated during the flow, even if not present at the microscopic level. This applies, in particular to the diverging coefficient $\sim \alpha^{-1}$ in front of the gauge fixing term. The value $\alpha=0$ is a (partial) fixed point to which the flow may be attracted in the infrared. 

The generation of a suitable gauge fixing term is sufficient for a realization of a local gauge symmetry for the effective action for the light modes. Local gauge symmetries can therefore arise in a rather general context. This does not imply, however, that all such gauge symmetric effective actions belong to the same universality class as standard Yang-Mills theories or quantum gravity. While local gauge symmetry is a crucial ingredient, it is presumably not sufficient for the determination of the universality class. In addition, $\G(g)$ should have suitable locality properties, for example admitting a derivative expansion on scales where perturbation theory applies (sufficiently above the confinement scale). Also the generalized measure contributions (Faddeev-Popov determinant or associated ghost sector) should be present.

Having established the way how local gauge symmetries can arise dynamically, the focus will have to concentrate on the properties of universality classes in order to find out under which circumstances the known fundamental interactions could emerge as a ``long-distance-property''.

\bigskip\noindent
{\em Acknowledgment}: This work is supported by ERC-AdG-290623.

\bibliography{Gauge_symmetry_from_decoupling}

\end{document}